\documentclass[sigconf]{acmart}
\renewcommand\footnotetextcopyrightpermission[1]{} 
\usepackage{stfloats} 
\usepackage{subcaption}
\usepackage{tcolorbox}   
\usepackage{layouts}
\usepackage{adjustbox}
\usepackage{mdframed}
\usepackage{float}
\usepackage{listings}
\usepackage{tikz}
\usepackage{comment}
\usepackage{xspace}
\usepackage{algpseudocode}
\usepackage{color}
\usepackage{colortbl}
\usepackage{enumitem}
\usepackage{balance}
\usepackage{soul}
\usepackage{multirow}
\usepackage{xcolor,pifont}
\usepackage{fancyhdr}
\usepackage[inkscapelatex=false]{svg}
\newcommand{\ballnumber}[1]{\tikz[baseline=(myanchor.base)] \node[circle,fill=.,inner sep=1pt] (myanchor) {\color{-.}\bfseries\footnotesize #1};}

\AtBeginDocument{%
  }

\begin{document}

\title{Empowering Vector Architectures for ML:\\ The CAMP Architecture for Matrix Multiplication\\ }

\author{
\begin{tabular}{c}
Mohammadreza Esmali Nojehdeh, 
Hossein Mokhtarnia, 
Julian Pavon Rivera,\\
Narcis Rodas Quiroga, 
Roger Figueras Bagué, 
Enrico Reggiani, 
Miquel Moreto,\\ 
Osman Unsal, 
Adrian Cristal, 
Eduard Ayguade\\
\textit{Barcelona Supercomputing Center*}
\end{tabular}
}





\begin{abstract}
This study presents the Cartesian Accumulative Matrix Pipeline (CAMP) architecture, a novel approach designed to enhance matrix multiplication in Vector Architectures (VAs) and Single Instruction Multiple Data (SIMD) units. CAMP improves processing efficiency of Quantized Neural Networks (QNNs).

Matrix multiplication is a cornerstone of machine learning applications, and its quantized versions are increasingly popular for more efficient operations. Unfortunately, existing VAs and SIMD-support units struggle to efficiently handle these quantized formats. In this work, we propose CAMP, a simple yet effective architecture that leverages a hybrid multiplier. The CAMP architecture significantly advances the performance of vector architectures in handling quantized data, enabling more efficient execution of matrix multiplication across various platforms, specifically targeting the ARMv8 Scalable Vector Extension (SVE) and edge RISC-V with SIMD-support unit architecture. In addition to increasing throughput, CAMP's architectural design also contributes to energy efficiency, making it an effective solution for low-power applications.

Evaluations on a range of Large Language Models (LLMs) and Convolutional Neural Networks (CNNs) demonstrate that matrix multiplication operations using the proposed micro-architecture achieve up to 17$\times$ and 23$\times$ performance improvements compared to their respective baselines, the ARM A64FX core and a RISC-V-based edge System-on-Chip (SoC). Furthermore, synthesis and place-and-route (PnR) of the CAMP micro-architecture using Synopsys tools—targeting ARM TSMC 7nm for A64FX and GlobalFoundries 22nm for the RISC-V SoC—add only 1\% and 4\% area overhead, respectively, compared to the baseline designs.

\end{abstract}
\thanks{*All authors are affiliated with the Barcelona Supercomputing Center. Corresponding author: Mohammadreza Esmali Nojehdeh (mohammadreza.esmali@bsc.es).}
\maketitle
\pagestyle{plain}

\section{Introduction} \label{Sec:Intro}

The remarkable resurgence of artificial intelligence (AI) in a variety of applications, from basic handwritten digit recognition~\cite{lecun2010mnist} to sophisticated chatbots~\cite{chatbot}, has significantly boosted both research and industrial interest in this domain~\cite{nature}. However, these advances in machine learning, especially in deep neural networks (DNNs), have led to a substantial increase in the demand for computational hardware resources for operation. Traditionally, the training of DNNs is conducted on high-performance architectures, after which the trained models are deployed onto relatively low-performance devices for inference purposes. Specifically, emerging large language models, such as ChatGPT, cannot operate efficiently on personal computers with single or double precision floating point numbers due to their immense computational requirements. These models necessitate vast amounts of memory and processing power, which are typically beyond the capabilities of standard consumer hardware, thereby requiring specialized or cloud-based solutions for both training and inference\cite{large_language_model}.

Quantization is a widely used technique to enable resource-limited devices to efficiently operate DNNs \cite{Enrico_HPCA, AQ2_micro}. In most state-of-the-art implementations, this method involves converting data representations from floating-point to integer formats, which significantly reduces memory usage and computational demands. By maintaining accuracy within a reasonable range, this approach enables the practical deployment of complex DNNs on resource-constrained devices.
In the realm of deep neural network quantization, several well-established techniques and frameworks have emerged, including TensorFlow Lite Quantization for mobile and edge devices~\cite{tensorflow}, NVIDIA's TensorRT for high-performance inference~\cite{NVIDIA}, Facebook's QNNPACK for optimized Convolutional Neural Networks (CNNs) operations~\cite{Facebook}, Google's gemmlowp for low-precision arithmetic in mobile apps~\cite{gemmlowp}, among others.

On the hardware side, different architectures have been proposed for training and inference in DNNs. These include custom Application Specific Integrated Circuit (ASIC) accelerators proposed in academia \cite{ecnn, dadiannao, envision, Micro_accelerator, Stellar_micro, Unico_micro, Diva_micro}, and many others in industry such as Google's tensor processing unit (TPUv5p)~\cite{google_cloud_tpu}, Apple neural engine M3~\cite{apple2023m3}, Amazon AWS inferentia2~\cite{aws_inferentia2} and Tesla Dojo chips~\cite{tesla2023dojowhitepaper}. Also Field Programmable Gate Arrays (FPGAs) like Xilinx Alveo Series \cite{xilinx_alveo}, and Intel Stratix series~\cite{Intel_FPGA} are designed for machine learning tasks.  GPUs like intel $X^e-HPG$ architecture~\cite{IntelXeHPG2023}, NVIDIA RTX Series~\cite{NvidiaAmpere2023}, Blackwell platforms~\cite{nvidia2024blackwell}, and AMD CDNA architecture for data centers~\cite{AMDCDNA32023} play a key role in both training and inference phase due to their high throughput.

In contrast to the competitive hardware design landscape for DNNs, Central Processing Units (CPUs) and Vector Accelerators (VAs) have relatively lagged behind, as they were not originally designed with deep learning applications in mind. Nevertheless, the ubiquity of CPUs in supercomputing centers and their flexibility continue to make them preferable for many workloads. Additionally, according to a study by Facebook~\cite{dnn_facebook}, CPUs demonstrate superior latency and flexibility when running DNNs alongside general-purpose tasks.

Only recently have newer generations of CPUs begun to incorporate dedicated systolic-array-based accelerators or co-processors for deep learning (DL), which, while effective, introduce significant area and design complexity due to their tightly coupled data movement and specialized compute patterns. Intel’s Deep Learning Boost technology~\cite{IntelDLBoost2023}, Vector Neural Network Instructions (VNNI)~\cite{intel2023}, and Arm’s Matrix Multiply Accumulate (MMLA) instructions~\cite{ARM_MMLA}, introduce specialized support for AI inference. However, they still lack dedicated integer units to efficiently handle low-precision data, which limits the potential benefits of quantization. As a result, although software frameworks continue to reduce precision to as low as 4 bits, insufficient hardware support hinders the full realization of these performance gains.

To illustrate this gap, Table~\ref{tbl:vec_arc} presents the performance of ARM and Intel vector units on 512×512 matrix multiplication, using 32-bit floating-point as a baseline and comparing against their respective 8-bit and 4-bit performance.

ARM SVE supports 8-bit integer multiplication, but it returns only the upper or lower 8 bits of the 16-bit result, requiring additional widening and reinterpretation operations. This overhead reduces the efficiency gains of quantization compared to FP32. To address this, Arm recently introduced the Scalable Matrix Extension (SME), which includes dedicated support for 8-bit operations with 32-bit accumulation.
The Apple M4 core, which is the only publicly available processor supporting SME, achieves a 2$\times$ speedup in 8-bit integer matrix multiplication compared to FP32, indicating that SME remains primarily FP32-centric~\cite{remke2024hello}.
 Intel’s Advanced Vector Extensions (AVX) with Integer Fused Multiply-Add (IFMA) instructions support 8-bit operations, achieving up to 4.5× better performance compared to FP32 operation on the Intel Xeon Platinum 8460Y+ (Sapphire Rapids), but still not support 4-bit operations.

Similarly, for low-power design, several studies based on SIMD units have been investigated to accelerate General Matrix Multiplication (GeMM) operations~\cite{pulp_mix,pulp,dory,cmix}. However, these studies are limited to small matrix sizes that fit within the L1 cache. Moreover, the need for additional instruction set architecture (ISA) support for sub bit-widths, along with the lack of consideration for the GeMM library structure, makes these approaches suboptimal and, in some cases, even worse than using higher bit-widths~\cite{pulp}.

To bridge this gap, we developed the Cartesian Accumulative Matrix Pipeline (CAMP) micro-architecture, designed to enhance integer multiplication operations within Arithmetic Logic Units (ALUs), rather than relying on complex solutions such as co-processors or accelerators. CAMP eliminates the need for specialized indexing modes or block divisions. By either replacing existing integer units or adding minimal area overhead, the CAMP architecture can be integrated into any vector or SIMD-style design. 

Inspired by highly optimized algorithms such as GotoBLAS~\cite{goto_high_performance}, our contribution centers on integrating intra-lane and inter-lane accumulators with hybrid multipliers to efficiently handle byte and sub-byte matrix multiplication. To ensure data locality and compatibility with GeMM libraries, we show that the outer product—referred to here as the Cartesian product\footnote{The Cartesian product is typically a concept from set theory, involving two sets $A$ and $B$, where $A \times B$ denotes the set of all ordered pairs $(a, b)$ with $a \in A$ and $b \in B$. In this work, we use the term "Cartesian" to refer to the operation in which all elements from one vector register are multiplied by all elements of another vector register, akin to the idea of Cartesian pairing but applied in a micro-architectural context. This usage emphasizes the comprehensive pairing of elements, similar to how the Cartesian product considers all possible pairs.}—is the essential operation in high performance libraries, in contrast to the inner product that underlies many prior designs~\cite{pulp,pulp_mix,ARM_MMLA}.
A key insight in CAMP's design is that the divide-and-conquer structure of the proposed hybrid multiplier aligns naturally with vector processing. This relationship between the outer-product computation model and the hybrid multiplier structure enables efficient execution across varying precision levels. Halving the operand bit-width doubles the number of vector elements, resulting in a fourfold increase in pairwise multiplications, perfectly matching the recursive nature of the hybrid multiplier. 

As shown in Table~\ref{tbl:vec_arc}, CAMP efficiently exploits quantization within vector ALUs, offering more than mere datatype conversion speedups by delivering performance gains through architectural efficiency. These results demonstrate that this lightweight, vector-integrated approach effectively bridges the performance gap for low-precision AI workloads without sacrificing general-purpose programmability.


\begin{table}[]
\caption{Speedup of Int8 and Int4 matrix multiplication over FP32 for square matrices of size 512 across architectures. (\ding{55} = supported via widening/packing, but excluded due to inefficiency and additional instruction overhead.)}

\centering
\scalebox{0.9}{ 
\begin{tabular}{cccc}
\hline
\textbf{Architecture} \hspace{18pt} & \textbf{FP32} & \textbf{Int8} & \textbf{Int4} \\ \hline
ARMv8+SVE \hspace{18pt} & x & \ding{55} & \ding{55} \\
ARMv9+SME \hspace{18pt} & x & 2x & \ding{55} \\
IntelAVX+IMFA \hspace{18pt} & x & 4.5x & \ding{55} \\
ARMv8+SVE/CAMP\hspace{18pt} & x & 7.4x & 12.4x \\
RISC-V/CAMP \hspace{18pt} & x & 14.1x & 25.1x \\
\hline
\end{tabular}
}
\label{tbl:vec_arc}

\end{table}

The primary contributions of this paper are outlined as follows:

\begin{itemize}
    
        


    \item \textbf{Custom Instruction Addition:} Introduction of custom instructions to ARM-SVE and RISC-V to expedite matrix multiplication operations, significantly enhancing GeMM performance. 

  \item \textbf{Alleviating Functional Unit Stall Rate:} Through the implementation of the CAMP architecture, the functional unit busy rate for matrix multiplication is reduced from 80\% to under 10\%, with functional unit stalls becoming negligible. This architectural innovation addresses previous compute limitations in VAs by optimizing arithmetic operations through efficient interlane accumulators and hybrid multipliers, effectively increasing the number of functional units. 

    \item \textbf{High Throughput and Energy-Efficient Design:} The proposed design achieves high throughput while maintaining energy efficiency. The RISC-V implementation reaches a throughput of 13 GOPS and 23 GOPS for 8-bit and 4-bit versions during convolution operations. Corresponding energy efficiency values are 270 GOPS/W and 405 GOPS/W for 8-bit and 4-bit operations, respectively.

    \item \textbf{Synthesis and Place and Route (PnR) of the Proposed Model:} Successful completion of synthesis and place and route processes for the RTL of the proposed architecture using ARM TSMC 7nm technology for the ARM core, and Global Foundries 22nm FDX technology for the RISC-V-based edge processor. The comprehensive design confirms the architecture's compactness, with the final layout occupying only 1\% and 4\% of the total core area compared to a single core of the ARM A64FX and RISC-V  System-on-Chip (SoC), respectively, demonstrating efficient area utilization without compromising performance.

    \item \textbf{Performance Evaluation in Transformers and Convolutional Neural Networks:} Evaluation of the proposed instruction across four LLM models (BERT base, BERT large, GPT-2 large, and GPT-3 small), four convolutional neural networks (AlexNet, VGG-16, ResNet-18, and MobileNet-v1), and square input matrix multiplication. Experimental results confirm the efficiency of the proposed instruction across all network layers. The proposed method achieves significant speedup, outperforming established libraries like OpenBLAS and gemmlowp by factors of \textbf{11}$\times$ and \textbf{6}$\times$ in average, respectively.

\end{itemize}

\section{Background and Motivation}
\label{Sec:Background}
Matrix multiplication is at the core of most deep learning applications. The following subsection explains the importance of this operation in Transformers and CNNs, as well as the existing libraries available for processing these computations.

\subsection{Transformers and Convolutional Neural Networks}

Large Language Models (LLMs), such as the Generative Pre-trained Transformer (GPT) leverage advanced transformer architectures to perform tasks in natural language processing. Notably, matrix multiplication operations account for over 80\% of the runtime in LLMs, underscoring their critical role in these models' performance~\cite{samsung}. Several methods have been proposed to accelerate transformers \cite{ham20203,wang2021spatten}. Similarly, CNNs are a specialized subset of machine learning algorithms. At their core are convolution layers that extract patterns from the input. The extracted patterns through convolution layers followed by other layers like pooling layers, normalizing layers or fully connected layers plays a key role in classifying and detection applications. A well-known technique to improve the execution time of convolution layers is to cast their operations as GeMM computations, which results in high performance computation, thanks to advanced matrix multiplication algorithms. The key function in this process is the image-to-column, or $im2col$, function. This method employs unrolling in a manner suitable for matrix multiplication, which is efficiently processed by highly optimized Basic Linear Algebra Subprograms (BLAS) libraries such as OpenBLAS~\cite{openblas}. This approach significantly speeds up computation, although it requires more memory~\cite{im2col}.

\subsection{BLAS Library}\label{Subec:ulmblas}
\begin{figure}
    \centering
    \includegraphics[width=\linewidth]{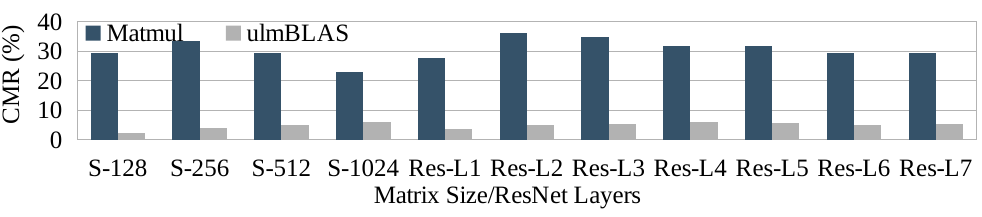}

    \caption{Cache Miss Rate (CMR) for matrix multiplication using naive and ulmBLAS methods on square matrices and ResNet layers, evaluated on the A64FX core.}

    \label{fig:cache_miss}
 
\end{figure}

To achieve near-optimal performance in implementing matrix multiplication in real-world applications, it is crucial to establish an optimized data flow. The optimization should reduce the memory footprint and ensure contiguous memory access. The naive implementation method, which accesses the first matrix in row-major order and the second matrix in column-major order, obtains the result by multiplying and accumulating the corresponding elements. However, this method is not optimal, especially for large matrices, where these matrices do not fit in the cache and registers. 
 To demonstrate this behavior, we evaluate the cache miss rate for the naive matrix multiplication. Experiments conducted on the Fujitsu A64FX processor demonstrate that cache  miss rate for level1 in naive implementation is considerably high. The dark columns in Fig.~\ref{fig:cache_miss} illustrate the cache miss rates for different Square Matrix Multiplication (SMM) sizes. Additionally, to reflect realistic scenarios, we cast ResNet layers into matrix multiplication form in our experiments.
 It can be observed that the cache miss rate for the naive implementation method ranges from 23\% to 36\%, which indicates the inefficiency of this approach. These results underscore the need for more efficient implementations aimed at improving memory locality.
Based on this principle, various approaches have been developed in mathematical libraries. BLAS are a group of highly optimized linear algebra solvers~\cite{BLAS}. 
BLIS library \cite{BLIS}, which is similar to BLAS, implements GeMM operations based on GotoBLAS \cite{Goto_anatomy, goto_high_performance}. Furthermore, several BLAS-like libraries have been published targeting different architectures. For instance, the Intel Math Kernel Library (MKL) \cite{intelMKL} is highly optimized for Intel CPUs and GPUs, while OpenBLAS \cite{openblas} is an open-source library extensively optimized for various hardware, including ARM SVE and AVX512.
 
Similarly, ulmBLAS \cite{ulmBLAS2021} represents a high-performance C++ implementation of BLAS, developed initially from a pure C implementation of BLAS functions. The development process consists of two phases: the first phase focuses on the foundational C implementation, while the second phase introduces optimizations such as loop unrolling, pipelining, and the integration of assembly code to replace the original C code. 

\begin{figure}[t!]
  \adjustbox{center}{
    \includegraphics[width=50mm]{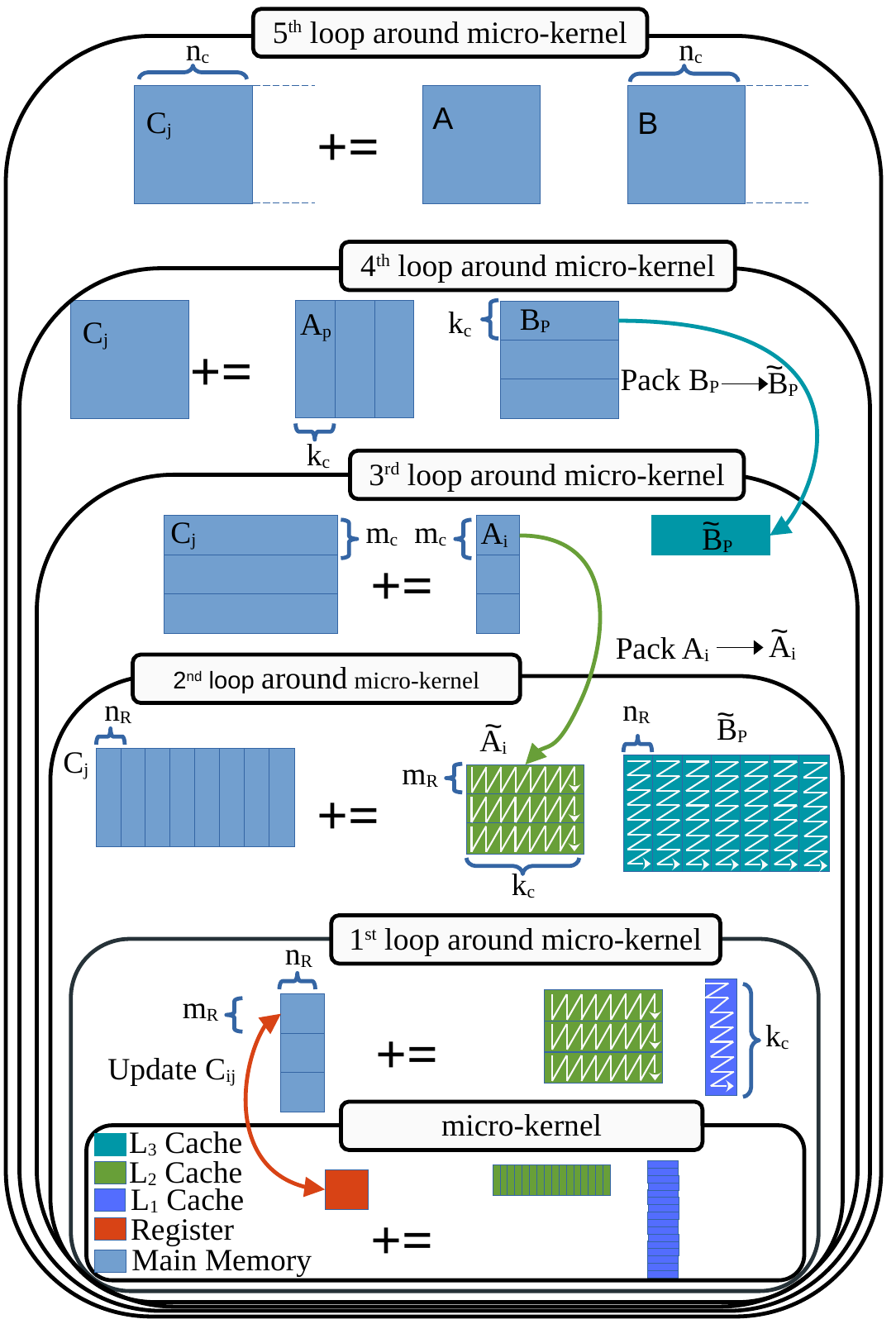}
  }

  \caption{The GotoBLAS algorithm for matrix-matrix multiplication\cite{van2017implementing}.}

  \label{fig:BLIS}
\end{figure}

GotoBLAS algorithm is the base for the GeMM operation in BLAS-like frameworks. This method systematically decomposes matrices into smaller blocks to maximize the ratio between computation and data movement. As depicted in Figure~\ref{fig:BLIS}, the method for blocking matrices across three cache levels (L1, L2, and L3) involves five nested loops around the micro-kernel to facilitate cache-coherent data movement. 
 
 

\begin{figure}[t!]
  \adjustbox{center}{
    \includegraphics[width=\linewidth]{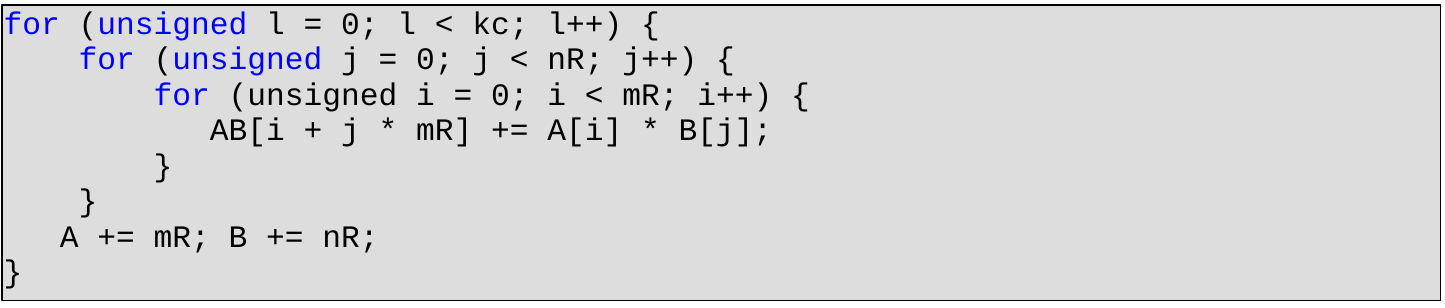}
  }

\caption{C++ Code for matrix multiplication using the GotoBLAS micro-kernel.}

  \label{fig_code:matrix_multiplication}

\end{figure}

The $5^{th}$ loop around micro-kernel decomposes matrices C and B into blocks with $n_c$ columns, referred to as panels. The $4^{th}$ 
loop further decomposes matrix B into panels with a height of $k_c$. These panels must fit within the L3 cache, and the value of $k_c$ is determined based on the target architecture. Accordingly, the $3^{rd}$ loop packs matrix A into panels with a width of $m_c$ and a height of $k_c$, where the value of $m_c$ is defined to ensure panel height aligns with the L2 cache size. The parameters $m_R$ and $n_R$ defines the width and height of the panels, accommodating the L1 cache and registers in the $2^{nd}$ and $1^{st}$ loops around the micro-kernel. 


 The C++ implementation of the micro-kernel, as shown in Figure~\ref{fig_code:matrix_multiplication} differs from the naive implementation (MATMUL). Unlike the naive method, this kernel accesses the first matrix operand in a column-major format and the second matrix operand in a row-major format. It is structured around three nested loops, each corresponding to a specific dimension of the input matrices. The algorithm incrementally computes the product of corresponding matrix elements, accumulating these products in a resultant matrix. Sequentially, it advances the data pointers in blocks, aligning them with the next segments of the matrices to be processed. This approach minimizes cache misses and enhances data locality, which is crucial for optimizing performance in large-scale matrix operations.

Accordingly, the GotoBLAS algorithm, regardless of the architecture or data type exploited is applicable in any scenario. The first phase of the ulmBLAS library, which includes only pure C code without any optimization, is considered as the implementation of GeMM operations in this study. 
To demonstrate the efficiency of the GotoBLAS algorithm, a matrix multiplication experiment was conducted on the Fujitsu A64FX processor, this time using UlmBLAS. As shown in Figure~\ref{fig:cache_miss} with light-colored bars, the cache miss rate dropped to under 5\% by employing ulmBLAS.

\subsection{Limitation of VAs for GeMM Operation}
\label{Subsec:Limitation}

\begin{figure}
  \adjustbox{center}{
    \includegraphics[width=\linewidth]{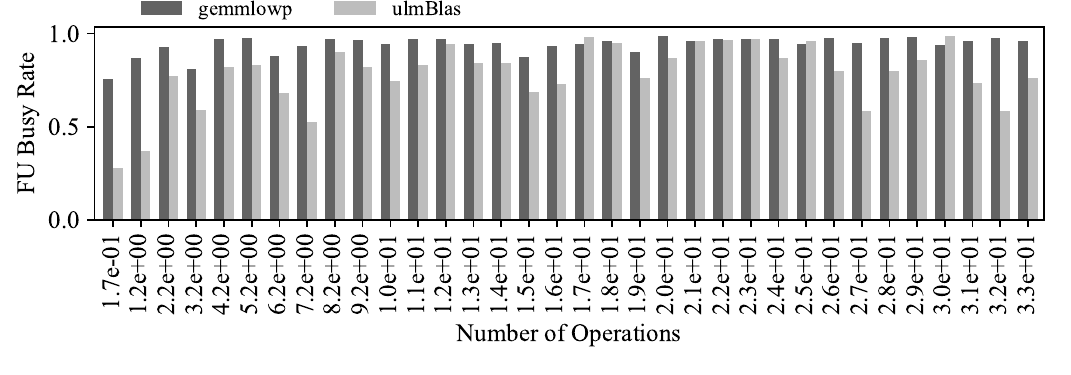}
  }

  \caption{Functional unit busy rate by method and number of operations.}
  \label{fig:functional_bar_busy_rate}

\end{figure}

VAs are typically implemented to address the demands of high-performance computing applications. By leveraging SIMD instructions, VAs are capable of executing multiple operations in parallel. 
The inherent data level parallelism processing capabilities of vector units significantly accelerate performance by reducing instruction count. 
On the other hand, the GotoBLAS algorithm efficiently improves the memory locality by splitting the input matrix into cache and vector register file friendly sub-matrices.
Since BLAS libraries and VAs are developed separately, the full potential of BLAS capabilities remains untapped for vector architecture ALUs. The primary observed limitation is as follows:

\textbf{Inadequate Number of Functional Units}:  While BLAS libraries improve data locality, limited functional units for quantized integer types can cause stalls. Multiply-and-add becomes a bottleneck in compute-intensive applications~\cite{madmac}. Evaluating GotoBLAS and gemmlowp on quantized CNNs with an A64FX system (Figure~\ref{fig:functional_bar_busy_rate}) shows functional unit busy rates exceeding 90\%, revealing hardware inadequacy.
 

\textbf{Neglecting Software in Instruction Development}
The previously proposed custom instruction focuses on inner products without considering highly optimized libraries. This approach results in inefficient load operations and introduces instruction overhead, particularly when dealing with byte and sub-byte data types. 
Considering software libraries and adding the capability to handle outer products can significantly reduce load operations and minimize unnecessary instruction overhead. 


\textbf{Challenges in Handling Integer Overflow in Vector Architectures}:
In floating-point arithmetic, overflow is handled within the same number of bits, making it straightforward to manage. However, this is not feasible for integer operations. For example, the product of two 8-bit integers results in a 16-bit value and the accumulation of these products in matrix multiplication can require up to 32 bits. This characteristic
conflicts with the principles of vector architectures, where input operands are vectors of numbers, and the output vector should maintain the same number of elements. 

The proposed CAMP micro-architecture addresses the outlined challenges. It performs matrix multiplication using a single instruction, minimizing data movement compared to the state of the art.  This approach simplifies the computational process and improves hardware resource utilization, significantly addressing the compute-bound nature of quantized matrix multiplication with highly optimized algorithms like the GotoBLAS algorithm in VAs and SIMD-supported units.


\section{Hybrid Multiplier}\label{Sec:hybd_mul}

\begin{figure}
    \centering
    \includegraphics [width=\linewidth]{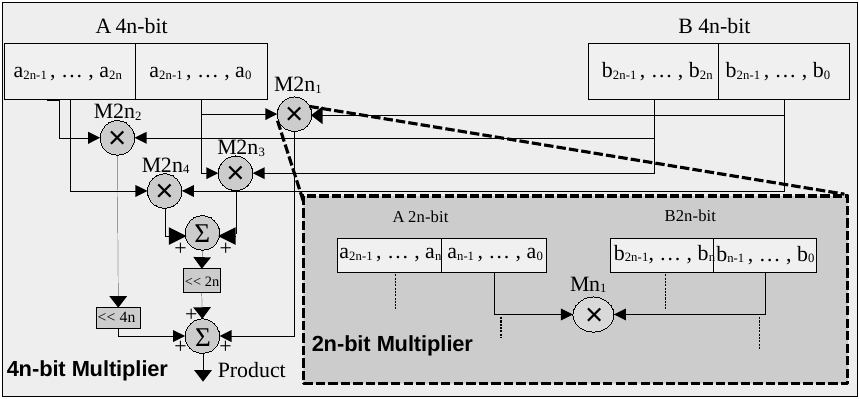} 

    \caption{Structure of hybrid multiplier for 4n-bit multiplication using n-bit building blocks.}

    \label{fig:hyb_str}

\end{figure}
 \begin{figure*}[ht]
    \centering

        \includegraphics[width=\textwidth]{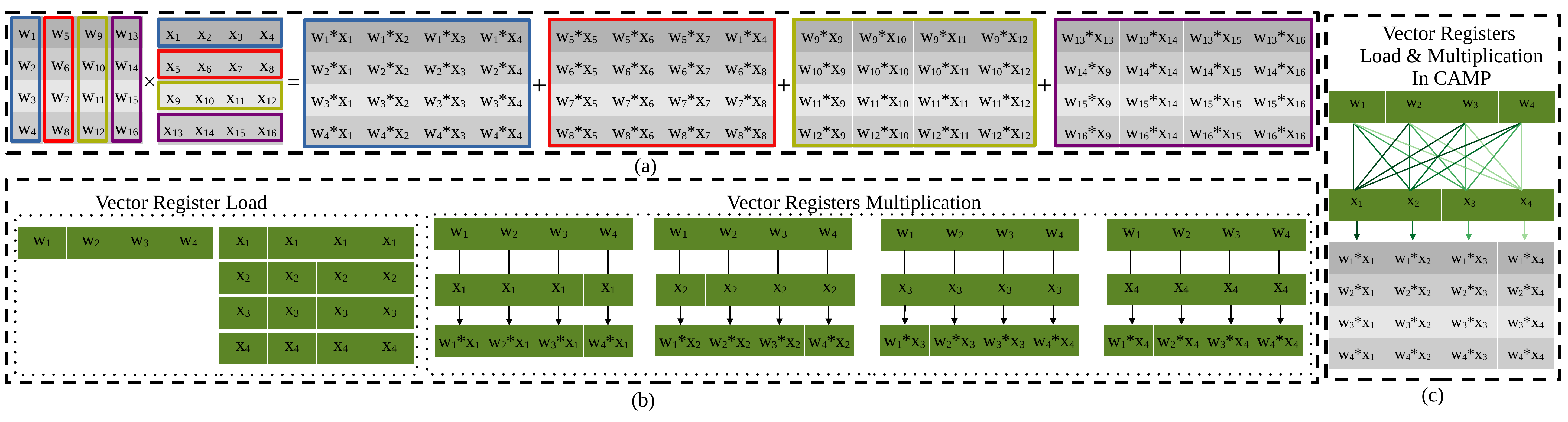}  

        \caption{ Matrix multiplication operation using the BLIS algorithm: (a) Illustrative representation of matrix multiplication, (b) Calculation of the first partial result with a generic vector processor, and (c) Calculation of the first partial result with the CAMP method.}
    
        \label{fig:mat_mul_3}

\end{figure*}

To address the identified challenges in outer product computations, we propose augmenting the vector processor with a dedicated block for integer operations.  Accordingly, we use a hybrid multiplier that leverages the principles of the 
divide-and-conquer approach proposed in~\cite{karatsuba} for multiplying large numbers. By hierarchically breaking down the multiplier into sub-multipliers of reduced bit-length, the hybrid multiplier supports multiplications across various bit lengths. This principle aligns with the concept of sub-byte operation support, where the substantial availability of sub-multipliers facilitates operations such as outer products, which require the deployment of extensive multipliers.
According to the algorithm, the multiplicand and multiplier of high bit-length multipliers are broken into two parts. For a 2N-bit multiplier and multiplicand, the multiplier division is represented as follows:


\begin{equation}
    \label{Eq:hybrid1}
    A = a_1 2^N + a_0 ,  \: B = b_1 2^N + b_0 
\end{equation}

The multiplication of $A$ and $B$ is calculated by:
\begin{equation}
    \label{Eq:hybrid3}
    P = a_1 \times b_1 2^{2N} + (a_1 \times b_0 + a_0 \times b_1) 2^N + a_0 \times b_0
\end{equation}

 

\begin{figure}
    \centering
    \includegraphics[scale=0.55]{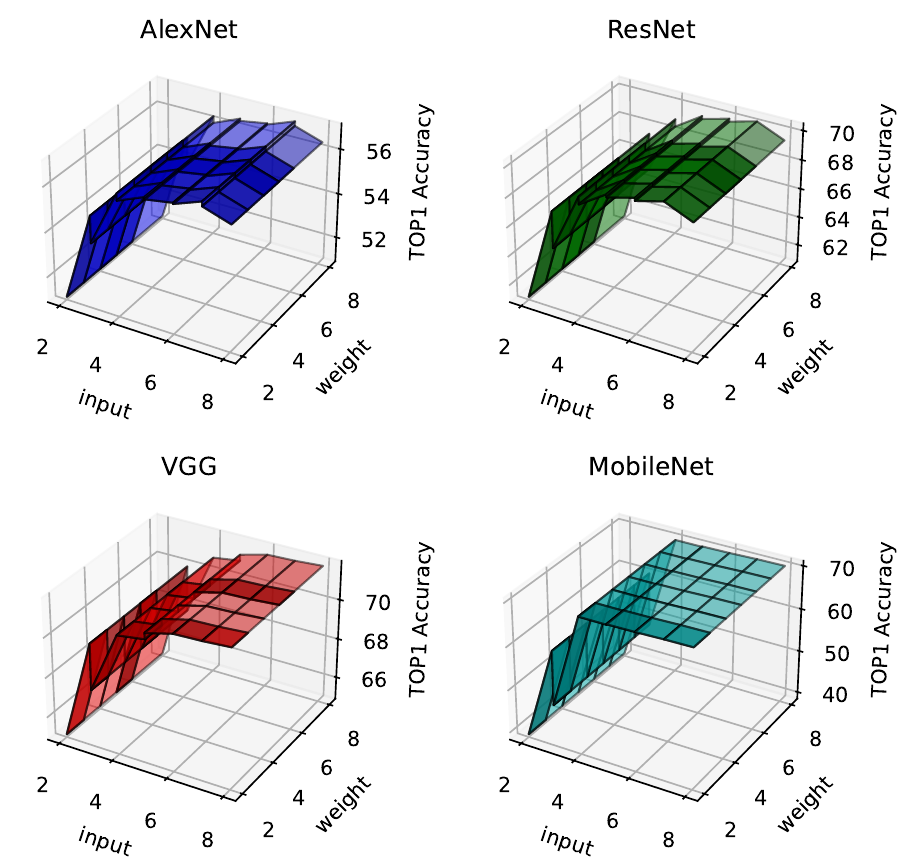}

    \caption{Relationship between weight precision and input bit-width with respect to model accuracy for various CNN architectures.}
    \label{fig:CNN_accuracy}

\end{figure}

According to the method, $N$-bit multipliers are decomposed into $N/2$-bit multipliers, a pattern that can be continued recursively until reaching 1-bit operands. The tiling design allows for selecting an optimal bit-length as a building block and placement of registers to align with the processor's maximum clock frequency and application-specific requirements.

To determine the optimal bit length for the hybrid multiplier, we refer to a survey examining the relationship between weight and input bit-width with respect to accuracy \cite{Enrico_HPCA}. According to that study and as shown in Figure~\ref{fig:CNN_accuracy}, the accuracy of different CNNs remains reasonable down to 4-bit quantization. However, reducing the bit-width below 4 bits leads to a significant degradation in accuracy.
In our study, which focuses on CNNs and LLMs, we set 4 bits as the minimum supported bit-width. However, depending on design requirements and designer intuition, the bit-width of the building block can be adjusted, allowing flexibility to optimize for specific performance and power targets.
In our study, 4-bit multipliers serve as the building blocks for constructing thirty-two 8-bit signed integer multipliers, where each 8-bit multiplier is composed of four 4-bit multipliers. Adders and shifters are integrated into these multipliers, as depicted in Figure \ref{fig:hyb_str}. By employing this approach, we not only enable the ALU to support multiple 8-bit multipliers for the outer product operation, but also add 4-bit support without requiring any instruction overhead for packing or unpacking data. The hierarchical nature of the divide-and-conquer algorithm allows it to function effectively as a multiplier for different bit lengths; however, in this study, we focus specifically on its capability to handle the outer product.

\section{CAMP Architecture}
\label{sec.cartesian}

\begin{figure*}[ht]
    \centering

        \includegraphics[width=\textwidth]{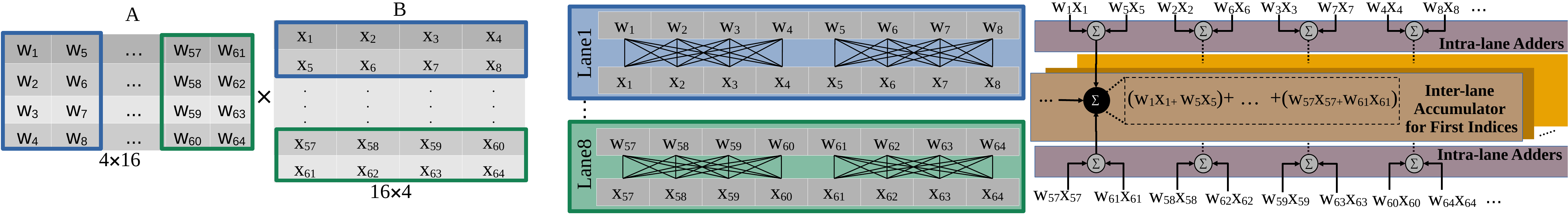}  

         \caption{CAMP matrix multiplication flow. Each lane computes outer products using hybrid multipliers and accumulates results via 16 intra-lane and 16 shared inter-lane accumulators. Lane 1 and Lane 8 are shown for clarity.}

        \label{fig:lanes_diagram}

\end{figure*}

Before delving into the details of our proposed architecture, we analyze the generic vector-based method and compare it with the proposed CAMP method to understand the importance of the outer product. 
In the first stage of this algorithm, the first column of the first matrix and the first row of the second matrix are selected. Their outer product, shown in blue in the figure, forms the first partial result of the matrix. Similarly, the outer products of subsequent columns of matrix A and rows of matrix B, highlighted in different colors, generate additional partial matrices. The final matrix is formed by accumulating partial results. 

Generic vector processors, as shown in Figure~\ref{fig:mat_mul_3}.b, only support element-wise operations. In this example, we neglect the data type and vector register length to simplify the explanation. Achieving the first partial matrix result requires loading the first column of the matrix into a vector register and duplicating the elements of the corresponding row across four different vector registers. This method not only increases the number of load operations 
but also places additional demand on the limited architectural vector registers. 
 Moreover, the presence of extra arithmetic instructions in the execution pipeline imposes additional overhead on the vector processor. This overhead arises due to increased instruction dependencies, higher register pressure, and potential stalls in the execution units, ultimately impacting performance and throughput.

To address these issues, the CAMP architecture, as shown in Figure~\ref{fig:mat_mul_3}.c, leverages the outer product operation instead of element-wise or inner-product-based computations. This approach requires only two load operations to fetch all the necessary data for the outer product, significantly reducing register usage and load operations.
Moreover, the enhanced multiplication mechanism, which supports different bit-widths, eliminates stalls caused by the limited functional units in traditional matrix multiplication.

In contrast to the generic method, where registers must retain their data throughout the accumulation process, the CAMP architecture incorporates an auxiliary register. This addition minimizes the number of load and store operations, improving efficiency.

According to the proposed CAMP architecture, matrix multiplication operates on input sizes of \textbf{4$\times$16} and \textbf{16$\times$4} for \textbf{8-bit} data, and \textbf{4$\times$32} and \textbf{32$\times$4} for \textbf{4-bit} data. The example shown in Figure~\ref{fig:lanes_diagram} illustrates the \textbf{8-bit} case, where each matrix fits entirely within a vector register.
The vector register length is \textbf{512 bits}, and the architecture is composed of \textbf{8 lanes}. As a result, each lane receives a \textbf{64-bit} segment of the vector register, corresponding to \textbf{8 elements of 8-bit data}. The input operands are sequentially distributed across lanes, with \textbf{Lane 1} receiving elements 1–8, \textbf{Lane 2} elements 9–16, continuing up to \textbf{Lane 8} with elements 57–64, enabling parallel computation.

The first operand (matrix A) is stored in \textit{column-major} order, and the second operand (matrix B) in \textit{row-major} order. Each lane performs an \textit{outer product} computation over two halves of its input data. For 8-bit operands, the first 4 elements from each operand form a 4$\times$4 outer product, requiring 16 8-bit multipliers. The second half repeats the same process, totaling 32 8-bit multipliers per lane.

If the input operands are 4-bit, each half contains 8 elements, forming an 8$\times$8 outer product, which requires 64 4-bit multipliers per half, and 128 4-bit multipliers in total per lane. This scaling is seamlessly supported by the hybrid multiplier design, where each 8-bit multiplier is internally composed of four 4-bit multipliers.

The hybrid multiplier architecture follows a recursive divide-and-conquer approach, where multiplying two $2n$-bit numbers requires four $n$-bit multipliers. This structural decomposition aligns naturally with the behavior of outer products in vector processors. Specifically, when the operand bit-width is halved, the number of elements in each vector register doubles, which in turn increases the number of pairwise multiplications in the outer product by a factor of 4. Therefore, the number of required multipliers grows proportionally with the reduction in bit-width, which matches the inherent scaling of the hybrid multiplier. This alignment ensures that sub-byte operations (e.g., 4-bit) can be efficiently supported within the vector processor pipeline without requiring fundamental changes to the execution model.

Within each lane, \textbf{16 intra-lane adders} are responsible for summing the corresponding multiplication results produced by the hybrid multipliers. Each adder accumulates a specific index across all outer product pairs computed within the lane. The outputs from all 8 lanes are then accumulated using a shared set of \textbf{16 inter-lane accumulators}, one per output index, to produce the final result matrix. 
\textit{For clarity, only Lane 1 and Lane 8 are shown with inter-lane accumulation paths in this figure.}

\subsection{$camp$ Instruction} \label{subsec:camp_isa}


The $\texttt{camp(VR0, VR1, VR2, mode)}$ instruction performs matrix computations within the CAMP architecture. The input operands (VR1, VR2) contain matrix values loaded from memory, while VR0 stores the result. The mode operand specifies the data bit-width.

\begin{figure}[t]
    \centering
    \includegraphics[width=\linewidth]{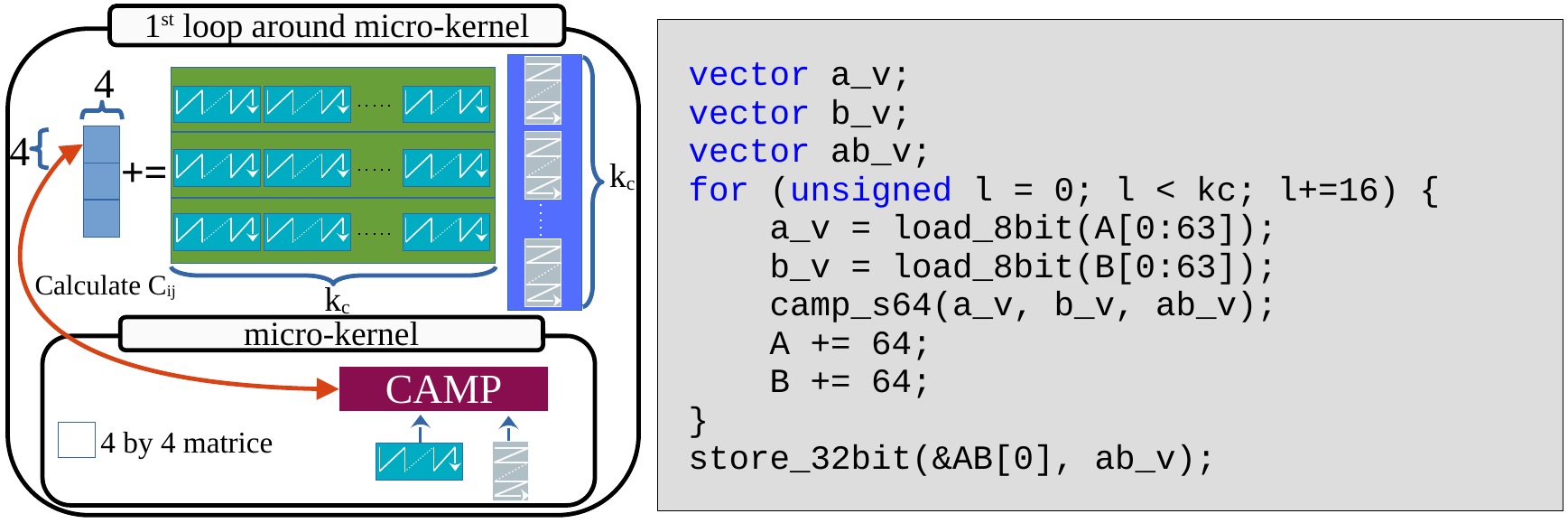}

    \caption{Integration of the CAMP micro-architecture with the GotoBLAS algorithm and the corresponding computational flow diagram.}
 
    \label{fig:combined_camp_blis}
\end{figure}

The code listing in Figure~\ref{fig:combined_camp_blis} illustrates the implementation of the \texttt{camp} instruction within the micro-kernel. Each call to \texttt{load\_8bit} fetches 64 elements of 8-bit data, filling a 512-bit vector register. The pointer increment \texttt{A += 64} advances by 64 bytes (512 bits) to the next vector-aligned chunk.
With $m_R = n_R = 4$, the innermost two loops in the GotoBLAS micro-kernel (Figure~\ref{fig_code:matrix_multiplication}) are eliminated, and the outer loop runs for $kc/16$ iterations. In each iteration, a $4 \times 16$ matrix from operand A is multiplied by a $16 \times 4$ matrix from operand B using the \texttt{camp} instruction, and the result is accumulated in an auxiliary register. After $kc/16$ iterations, the final result is stored in memory.

The GotoBLAS framework supplies the micro-kernel with blocks of size $4 \times kc$ and $kc \times 4$, which are processed incrementally through these $4 \times 16$ and $16 \times 4$ sub-blocks in each iteration of the loop.


This instruction also supports a 4-bit matrix multiplication operation. The inputs are two operands in vector register format, representing matrices of dimensions 4$\times$32 and 32$\times$4, each containing 4-bit data. The resulting output is a 4$\times$4 matrix of 32-bit elements, stored in the output vector register.

\subsection{CAMP Hardware} \label{subsec:Microarchitecture}
\begin{figure}[t]
    \centering
    \includegraphics[width=70mm]{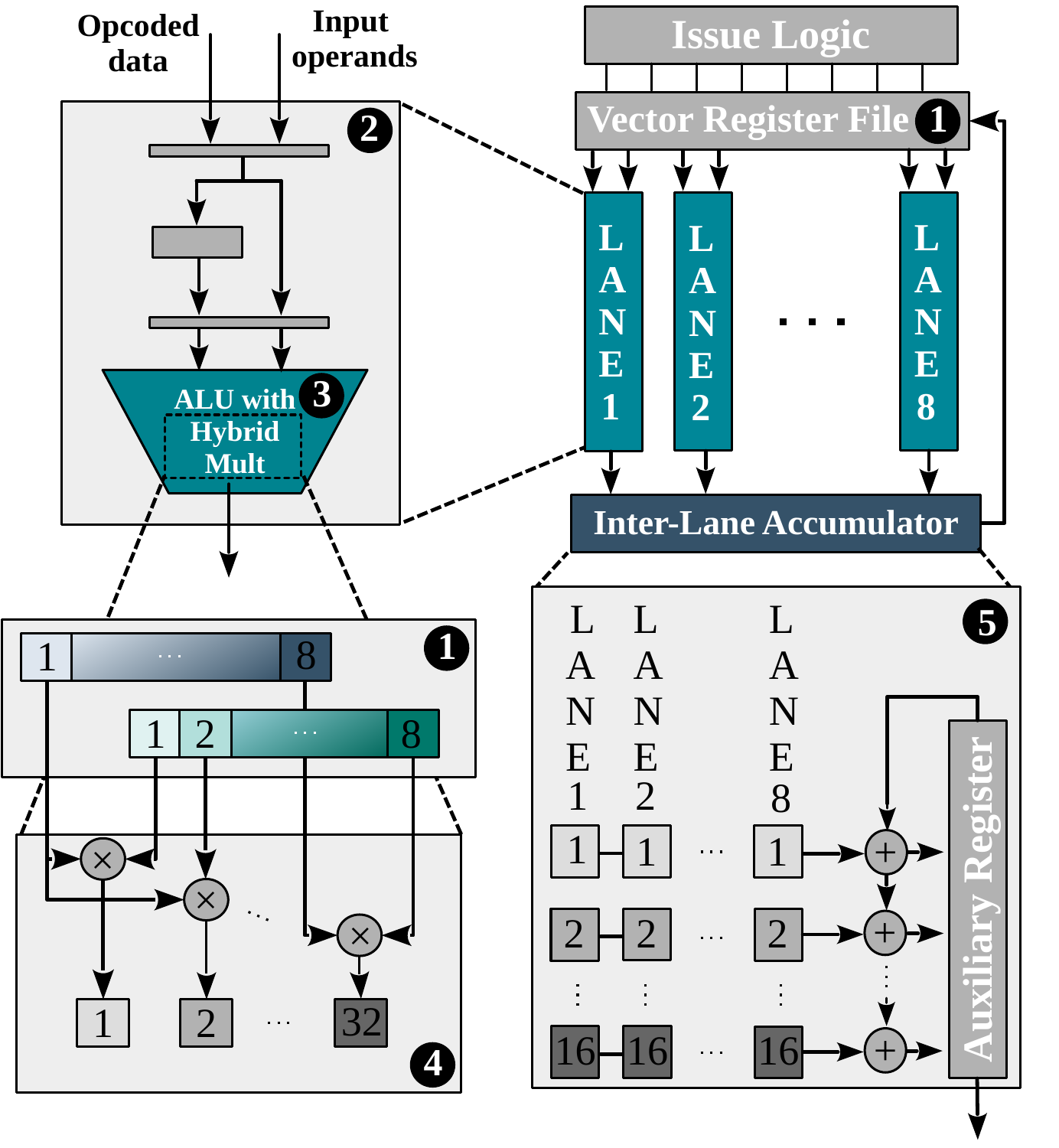}

    \caption{CAMP micro-architecture schematic.}

    \label{fig:hyb_VPU}
 
\end{figure}

In accordance with the proposal outlined in Section \ref{Subsec:Limitation}, we have integrated a hybrid multiplier as an additional functional unit within ALU. In this part we investigate the execution unit of the ARM architecture, extended with SVE capabilities, but it is applicable to any other VA or SIMD-support unit. 
Figure~\ref{fig:hyb_VPU} provides an overview of how the ALUs across different lanes are equipped with the hybrid multiplier. The vector register in Register File \ballnumber{1} has a length of 512 bits. To manage this data efficiently, the vector is divided into smaller segments, with each 64-bit segment allocated to a different lane for parallel processing. Each lane, as indicated by \ballnumber{2} operates independently. The ALUs, marked by \ballnumber{3} are outfitted with a hybrid multiplier. Based on the proposed micro-architecture, two 64-bit vectors fetched from \ballnumber{1} are passed into the hybrid multiplier \ballnumber{4}. This unit is comprised of 32 8-bit multipliers, enabling it to perform outer multiplication on two input operands. 

According to the GotoBLAS algorithm, to obtain the final result, we must still accumulate the results with the same index from different lanes. For instance, results with index 1 from all 8 lanes are accumulated to generate the result matrix's index 1. 
Similarly, the remaining 15 results from the lanes are accumulated with their corresponding indices in the other lanes by an Inter-Lane Accumulator \ballnumber{5}. It's worth noting that the registers in \ballnumber{5} for Lane 1 are the same as those in \ballnumber{4}, and this structure is consistent across the rest of the lanes. We have opted not to depict them in this figure for the sake of simplicity.


\subsection{CAMP Compilation} \label{subsec:camp_compile}
The CAMP architecture does not affect the front end or issue logic of the Vector Processing Unit (VPU). Instructions are fetched, decoded, and issued through the standard arithmetic path. Afterward, data is fed to the CAMP units within the lanes for the computation.
To expose the CAMP micro-architecture to the software stack, first, we extended the ARM SVE  ISA with the instruction introduced in Section~\ref{subsec:camp_isa}.
Second, we used vector intrinsic to include our instruction in the evaluated algorithms.
To support the CAMP micro-architecture within the RISC-V ISA, we added assembly inlining for the extended ISA to the target library. This enhancement ensures that the instruction is properly invoked for matrix multiplication operations, eliminating the need for programmers to interact with low-level assembly code. 


\section{Methodology}
\label{sec:methodology}

\subsection{Simulation Framework}
\label{sec:simulator}

To show the applicability of the proposed method, we evaluate the functionality and performance of the CAMP matrix multiplication on two different architecture ARM-SVE and RISC-V based edge SoC with SIMD-support unit.

For ARM vector architecture by using the gem5 simulator~\cite{gem5}
we simulate an ARM full-system running an Ubuntu 18.04 OS with a 4.9.4 Linux kernel.
We modeled and validated gem5 against a Fujitsu A64FX CPU~\cite{Fugako}.
Table~\ref{table:gem5_config} summarizes the main simulation parameters.
We extend the VPU model in gem5 to include the functionality of the hardware structures described in Section~\ref{subsec:Microarchitecture}.
To evaluate the performance of the CAMP micro-architecture in RISC-V, we integrated its hardware into an edge RISC-V SoC \cite{sargantana}. The target edge processor, implementing the RV64G instruction set, and a subset of the vector instruction features a single-core, 7-stage, in-order, single-issue pipeline, while the memory hierarchy includes L1 and L2 data caches with sizes of 32 KB and 512 KB, respectively. We compiled the GeMM operation with the RISC-V GNU compiler toolchain \cite{riscv_gnu_toolchain} extended with the proposed custom instructions, and executed bare-metal simulation.

\begin{table}[t!]
\begin{center}
\footnotesize
\caption{Simulated Systems Setup.}

\label{table:gem5_config}
\vspace{-0.2cm}
\scalebox{0.86}{
\begin{tabular}[width=\linewidth]{ l }
\hline \\ [-1.5ex]
\textbf{CPU:} 2.0 GHz, 16 cores, A64FX-like~\cite{odajima2020preliminary, Fugako} superscalar OoO\\
\textbf{Vector ISA}: ARM SVE ISA - Vector Length: 512b\\
\hline \\ [-1.5ex]
\textbf{L1-I:} 64KB, 8-way assoc., load-to-use = 2 cycles, Stride prefetcher\\
\hline \\ [-1.5ex]
\textbf{L1-D:} 64KB, 8-way assoc., load-to-use = 4 cycles, Stride prefetcher\\
\hline \\ [-1.5ex]
\textbf{L2 Cache:} 8MB, shared, 16-way assoc., load-to-use = 37, Stride prefetcher\\
\hline \\ [-1.5ex]
\textbf{DRAM:} 4-channel HBM2 \\
\hline \\ [-1.5ex]
\end{tabular}
}
\end{center}

\end{table}

\subsection{Benchmarks}
\label{sec:benchmarks}

We evaluate the efficiency of CAMP using BERT base, BERT large, GPT-2 large, and GPT-3 small as benchmarks. The matrix multiplications in the self-attention and feed-forward layers are specifically considered for this evaluation, focusing solely on inference operations rather than training. Additionally, we assess the performance with four different CNN models: ALEXNET, VGG, MOBILENET, and RESNET. The convolutional layers in these CNN models are cast into matrix multiplications. 
Table~\ref{tab:transposed_benchmark_sizes} reports the matrix sizes evaluated for each benchmark.





\begin{table}[]
\caption{Parameter Values in mk $\times$ kn Matrix Multiplication in Different Layers of Benchmarks.}

\scriptsize
\setlength{\tabcolsep}{2pt}
\scalebox{0.93}{
\begin{tabular}{cccccccccccccccccc}
\hline
\textbf{Size Index}        & \textbf{ALEXNET}      & \textbf{SMM}            & \textbf{RESNET}    & \textbf{VGG}           & \textbf{MOBILENET}        \\ 
                           & m,n,k                 & m,n,k                  & m,n,k              & m,n,k                  & m,n,k                     \\ \hline
\textbf{1}                 & 169,256,3456          & 32,32,32               & 12544,64,147       & 12544,128,1152         & 2544,32,27                \\
\textbf{2}                 & 169,384,2304          & 64,64,64               & 196,256,1152       & 12544,128,576          & 12544,64,32               \\
\textbf{3}                 & 169,384,3456          & 128,128,128            & 196,256,2304       & 196,512,4608           & 196,512,256               \\
\textbf{4}                 & 3025,96,363           & 256,256,256            & 3136,64,576        & 3136,256,1152          & 196,512,512               \\
\textbf{5}                 & 729,256,2400          & 512,512,512            & 49,512,2304        & 3136,256,2304          & 3136,128,128              \\
\textbf{6}                 & -                     & 1024,1024,1024         & 49,512,4608        & 50176,64,27            & 3136,128,64               \\
\textbf{7}                 & -                     & -                      & 784,128,1152       & 50176,64,576           & 49,1024,1024              \\
\textbf{8}                 & -                     & -                      & 784,128,576        & 784,512,2304           & 49,1024,512               \\
\textbf{9}                 & -                     & -                      & -                  & 784,512,4608           & 784,256,128               \\
\textbf{10}                & -                     & -                      & -                  & -                      & 784,256,256               \\
\hline 
\end{tabular}}
\label{tab:transposed_benchmark_sizes}

\end{table}

\subsection{Experiments}
\label{sec:experiments}







We compared the performance of CAMP using ulmBLAS against ARM-SVE running the optimized libraries listed below.


\textbf{1. Handv-int32:}
We implemented an in-house vectorized ulmBLAS implementation using the 32-bit integer SVE ISA.

\textbf{2. Handv-int8:}
We implemented a quantized 8-bit version of the algorithm to evaluate speedup from data-type conversion in vector architectures. Overflow handling was omitted, which may lead to incorrect results.

\textbf{3. OpenBLAS SGEMM:}
We directly used OpenBLAS's optimized SGEMM~\cite{openblas}.

\textbf{4. gemmlowp:}
This is the Google's GeMM vectorized implementation for low-precision data.

All the hand coded implementations have been vectorized using vector intrinsic.
Both OpenBLAS and gemmlowp are highly optimized using assembly, we directly used the open-source code unmodified.







\section{EVALUATION} \label{Sec:PERFORMANCE_EVALUATION}

\subsection{Physical Design}

To quantify the impact in the area and power, we  implemented the CAMP micro-architecture using Verilog and synthesized it using the $7\mathrm{nm}$ technology node, which is the same technology used by the ARM A64FX, with a target frequency of 2GHz.
We compare the area overhead of the CAMP to an A64FX core using the methodology proposed in~\cite{arima2021power}.

\begin{figure*}[ht]
    \centering
    \includegraphics[width=\textwidth]{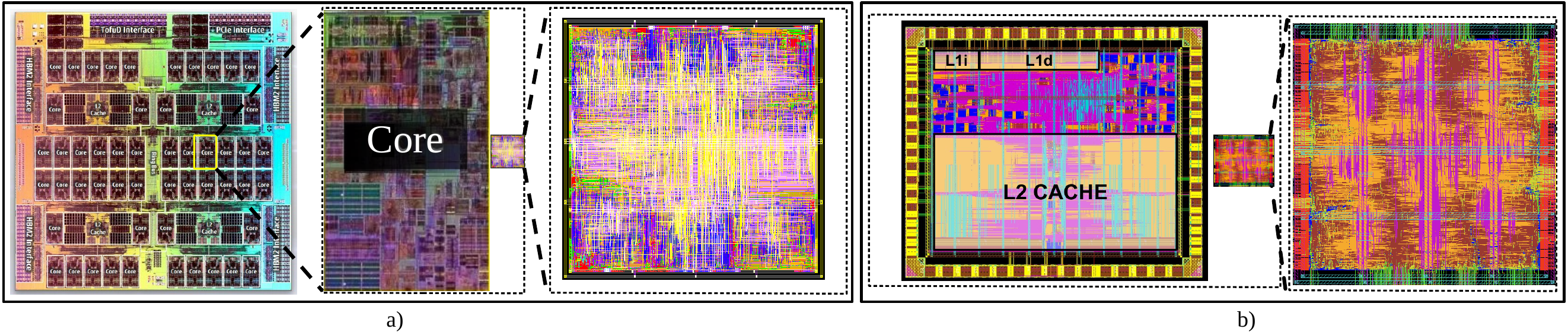}

    \caption{Area comparison between baselines and the CAMP micro-architecture: (a) Die photo of the A64FX CPU, showcasing its overall architecture and comparing the CAMP micro-architecture to a single core from the A64FX. (b) Floorplan of Sargantana SoC, compared to the CAMP micro-architecture.}
    \label{fig:layout}

\end{figure*}

The SoC layout and its relative areas, compared to the A64FX CPU and its core, and RISC-V SoC, are depicted in Figure~\ref{fig:layout}. The CAMP layout shown use a square floorplan, aiming to maximize cell density (around 85\% ) without causing congestion issues. Both  designs achieved positive slack time, ensuring that they do not impact the critical path.
Compared to the A64FX core, CAMP features a negligible area overhead of 1\%, with an area occupation of 0.027263~$\text{mm}^2$, utilizing Synopsys' ICC2 Place and Route tool \cite{synopsys}.
Similarly, we synthesized the CAMP micro-architecture using GlobalFoundries 22nm FDX technology for the RISC-V SoC, which is the same technology used by the baseline architecture \cite{sargantana}, targeting a frequency of 1 GHz. According to the experimental results, the CAMP micro-architecture occupies a total area of 0.0782 $mm^2$, adding an overhead of 4\% to the total chip area.

\subsection{Performance}

In this section, we analyze the performance of the CAMP across various LLMs and  CNNs, comparing it to the approaches introduced in Section~\ref{sec:experiments}.
 The performance metrics include the number of clock cycles and instructions, measured using the gem5 simulator for the ARM SVE architecture and bare-metal simulation for the RISC-V core, to assess the efficiency of the proposed micro-architecture.
By adopting the OpenBLAS approach on targeted A64FX core as a baseline, we normalize the clock cycle count relative to this approach for gem5 simulation results.
Furthermore, we consider the ratio of instruction numbers, expressed as a percentage, compared to the OpenBLAS approach as another metric for assessing the efficiency of the CAMP hardware. 
Since the instruction counts for handv-int8 and handv-int32 are similar, we report only the handv int-8 instruction count. Also, BLIS library by supporting 32-bit integer on edge RISC SoC, considered as the Baseline for SIMD-support units. 

\begin{figure}[tp]
    \centering
    \includegraphics[width=\linewidth]{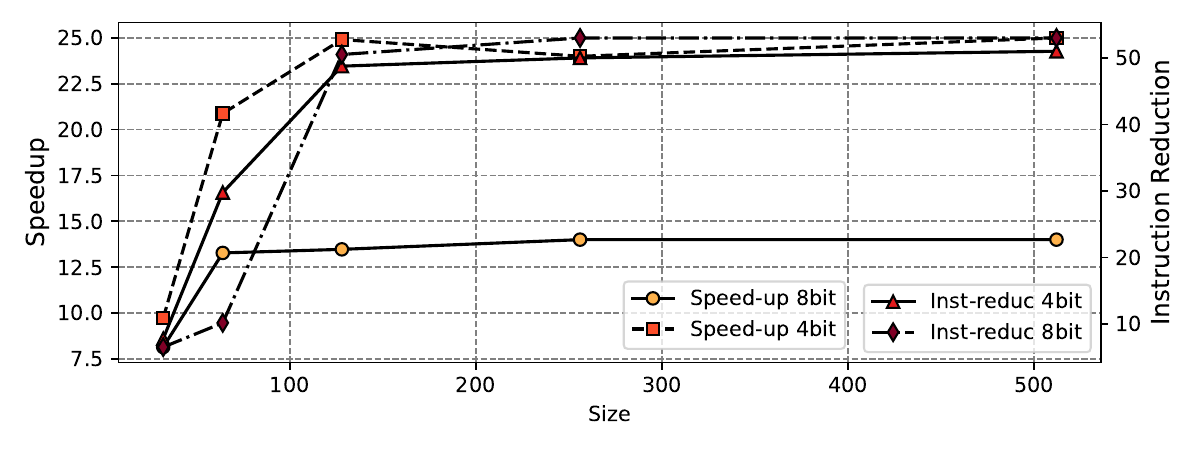}

    \caption{Comparison of normalized speed-up and instruction reduction times for square matrix multiplication using the Edge RISC-V core\cite{sargantana} baseline.}
    \label{fig:riscv_speedup}

\end{figure}

\sloppy
Figure~\ref{fig:riscv_speedup} illustrates the performance of the CAMP micro-architecture on a RISC-V SoC for SMM. The proposed method outperforms the baseline 32-bit BLIS library by approximately 24$\times$ in terms of clock cycles. This improvement is supported by the reduction in the number of instructions required for the CAMP micro-architecture compared to the baseline, highlighting its efficiency. Moreover, the absence of packing and unpacking instructions in the CAMP architecture leads to a linear relationship between 4-bit and 8-bit matrix multiplication. The CAMP micro-architecture achieves data throughputs of 16 GOPs and 28 GOPs for 8-bit and 4-bit operations, respectively. Additionally, according to experimental results, the proposed instruction achieves an energy efficiency of 270 GOPs/W and 405 GOPs/W for SMM with 8-bit and 4-bit operations, respectively.

The gem5 simulations provide a comprehensive analysis of the performance of the proposed micro-architecture in the targeted VA.

\begin{figure*}[t]
    \centering

        \includegraphics[width=\textwidth]{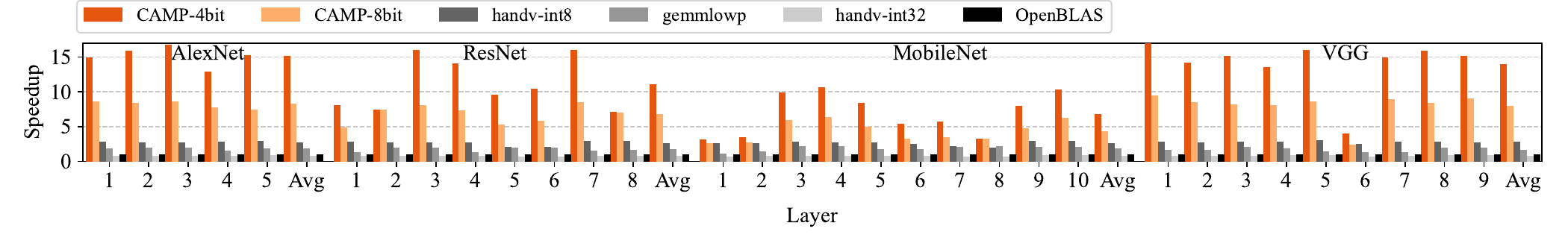}  

        \includegraphics[width=\textwidth]{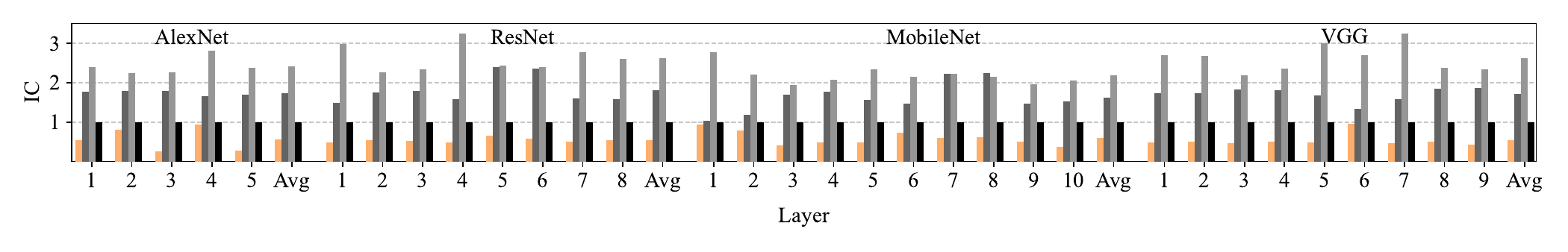}  

    \caption{Comparison of Normalized Speedup (Higher is Better) and Instruction Count (Lower is Better) for CNN Architectures,
using OpenBLAS on A64FX Core as Baseline. The ``Avg'' represents the average across all layers for each benchmark.}

    \label{fig:cnn_speedup}
\end{figure*}

\begin{figure}[t]  
    \centering
        \begin{subfigure}[t]{\columnwidth}
            \centering
            \includegraphics[width=\columnwidth]{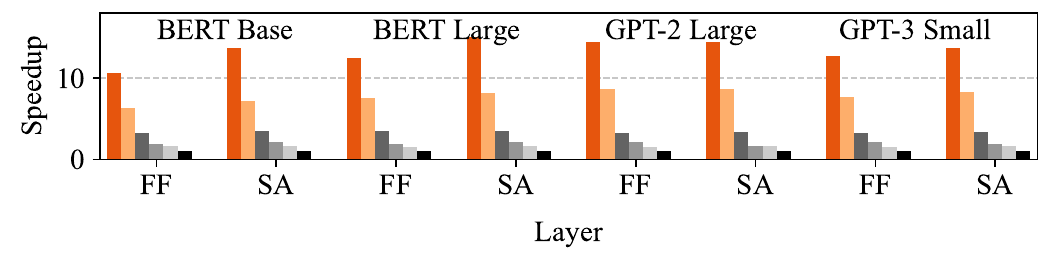}  
            \label{fig:speedup}
        \end{subfigure}

        \begin{subfigure}[t]{\columnwidth}
            \centering
            \includegraphics[width=\columnwidth]{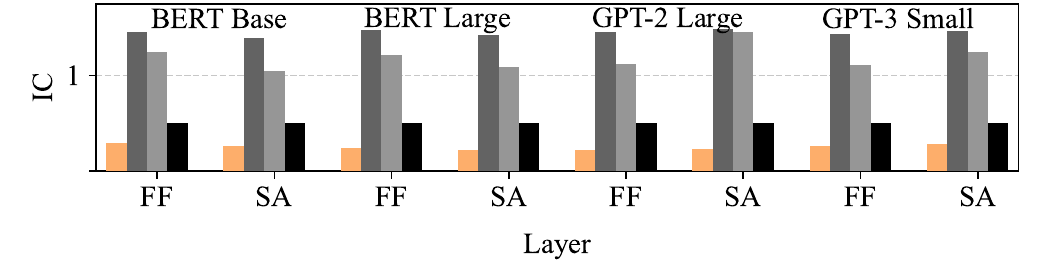}  
            \label{fig:instruction}
        \end{subfigure}

    \caption{Comparative analysis of normalized speedup (higher is better) and instruction count (lower is better) for LLMs, relative to OpenBLAS as baseline. FF denotes feed-forward layer, and SA denotes self-attention layer. The color coding for the benchmarks follows the legend provided in Figure~\ref{fig:cnn_speedup}.}

    \label{fig:llm_speedup}
\end{figure}

\begin{figure*}[t]
  \centering
  \includegraphics[width=\textwidth]{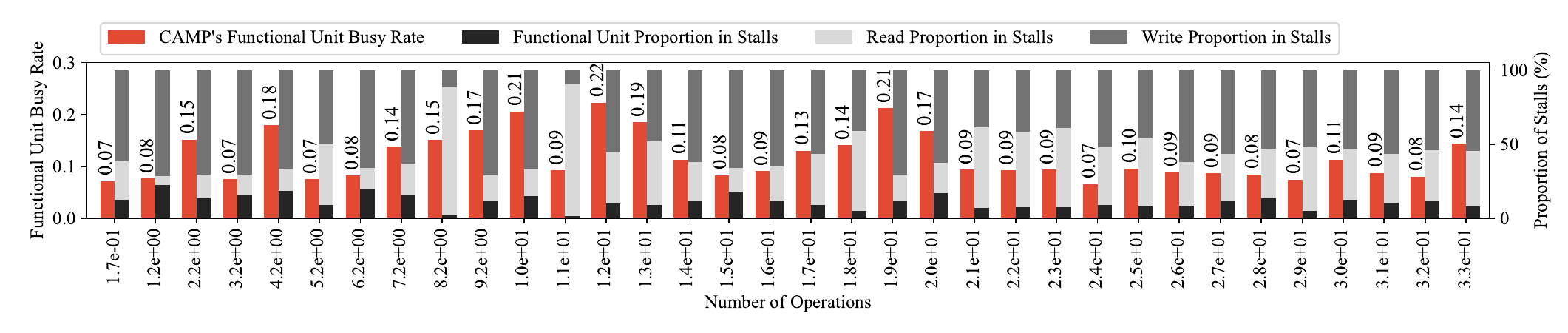}

  \caption{Comparison of CAMP functional unit busy rate and proportion of stalls by operation type.}
  \label{fig:camp_busy_rate}

\end{figure*}

\begin{figure}[t]
    \centering
    \includegraphics[width=\linewidth]{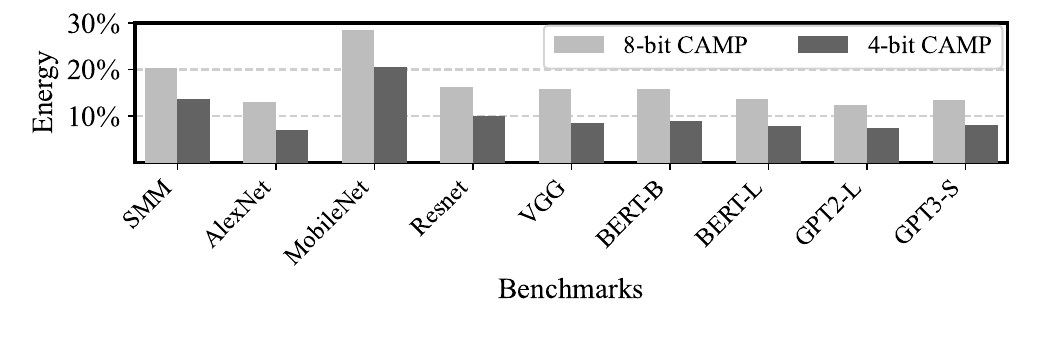}

    \caption {Normalized energy consumption of 8-bit and 4-bit CAMP implementations across different benchmarks. Energy values are expressed as a percentage relative to the baseline A64FX core (100\%).}

    \label{fig:energy}
\end{figure}

The most valuable data from this analysis is presented in Figures~\ref{fig:cnn_speedup}, \ref{fig:llm_speedup}, \ref{fig:energy}, and \ref{fig:camp_busy_rate}. Figure~\ref{fig:cnn_speedup} illustrates the performance results for all evaluated algorithms across various CNN layers, which the convolution layers are cast to GeMM computation, the size of the matrices for the evaluated CNNs are given in Table~\ref{tab:transposed_benchmark_sizes}. Additionally, Figure~\ref{fig:llm_speedup} represents the performance results for LLMs in self-attention and feed-forward layers.
The energy consumption of the CAMP extension required to perform all matrix multiplication operations for the investigated networks is shown in Figure~\ref{fig:energy}. 
 Energy consumption is quantified as the product of latency and power dissipation. The switching activity data, required for power dissipation analysis, was obtained from simulation using test data.
The energy values are normalized to the A64FX baseline core.

\begin{figure}[t]
    \centering
    \includegraphics[width=\linewidth]{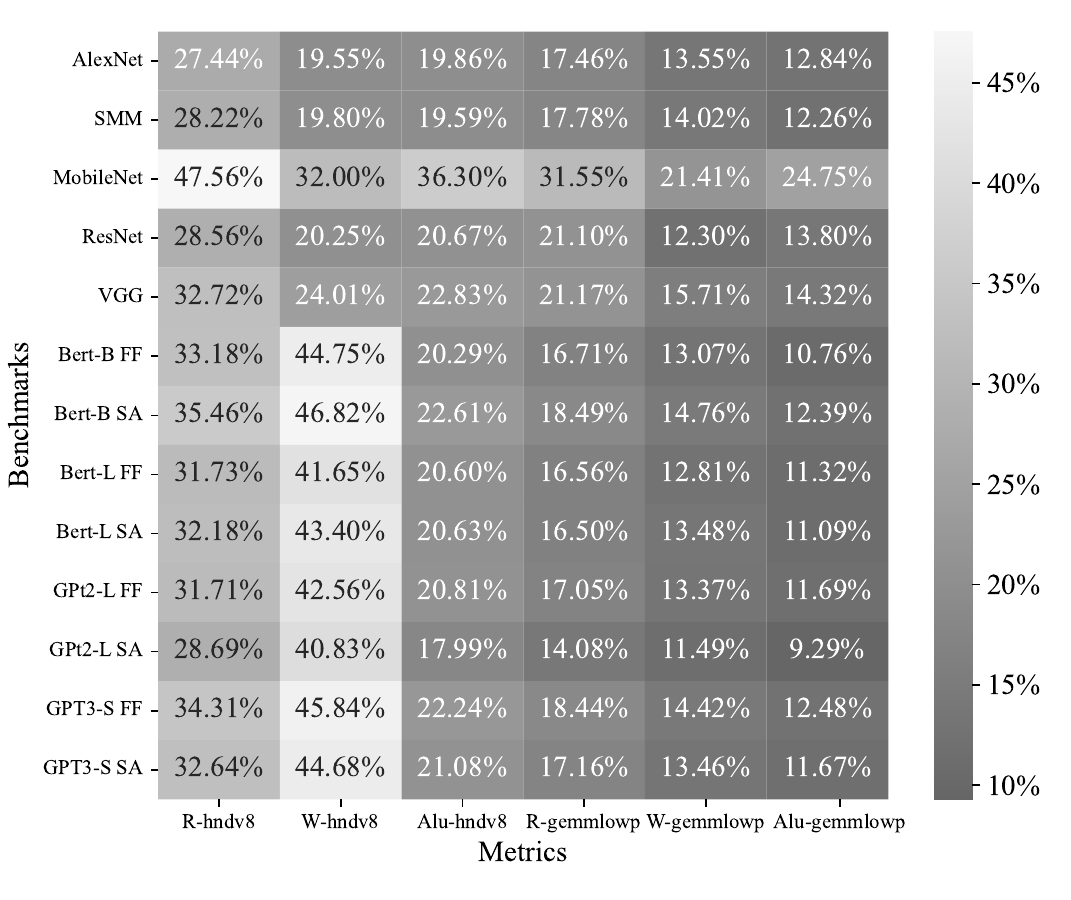}
    \caption {Heatmap showing the percentage of vector instructions used by the CAMP architecture compared to handv-int8  and gemmlowp implementations for CNN and LLM benchmarks. The x-axis represents instruction types (R for Read, W for Write, Alu for Arithmetic Logic Unit), and the y-axis lists benchmarks including AlexNet, SMM, MobileNet, ResNet, VGG, and various layers of BERT (Base, Large) and GPT (Large, Small) models. FF denotes Feedforward layer, and SA denotes Self-Attention layer (lower is better).}
    \label{fig:heat_map}
\end{figure}

Figure~\ref{fig:camp_busy_rate} illustrates the efficiency of the CAMP architecture through its busy rate and stall proportions. The bars represent the functional unit's busy rate, with stacked segments indicating stalls categorized as Functional Unit, Read, and Write across different operations.
Additionally, the heat-map in Figure~\ref{fig:heat_map} illustrates the percentage of vector instructions used by the CAMP architecture compared to handv-int8 and gemmlowp implementations across various CNNs and LLMs. Lower percentages indicate a reduction in the number of instructions, showcasing the efficiency of the CAMP architecture. We chose 8-bit implementations to ensure the same data volume for load and store operations in vector registers. The x-axis categorizes vector instructions for handv-int8 and gemmlowp implementations, and the y-axis lists the benchmarks.
%
%


Below are the key findings from the experimental results:


\textbf{Superior Performance of CAMP Architecture in Matrix Multiplications for LLM Transformers and CNNs}: 
The results in Figure~\ref{fig:cnn_speedup} and Figure~\ref{fig:llm_speedup} demonstrate that the proposed CAMP architecture—enabled through a software-hardware co-design approach using a customized ulmBLAS library—significantly outperforms baseline architectures running highly optimized software libraries such as OpenBLAS and gemmlowp. Notably, CAMP4-bit achieves up to 16$\times$, 11$\times$, 16$\times$, and 17$\times$ reductions in clock cycles across different layers of ALEXNET, MOBILENET, RESNET, and VGG, respectively, compared to OpenBLAS. It also outperforms gemmlowp, with improvements of 8$\times$, 5$\times$, 10$\times$, and 11$\times$ on the same networks.
For LLM workloads, CAMP4-bit achieves up to 15$\times$ speedup in clock cycles over OpenBLAS across various layers. Additionally, as shown in Figure~\ref{fig:energy}, CAMP reduces energy consumption by over 80\% relative to the A64FX core, while increasing peak power consumption by only 0.6\%. This tradeoff results in a performance gain of up to 17$\times$, highlighting the efficiency of our co-designed stack.

\textbf{Validating CAMP Efficiency Beyond Data Type Variations}:
To ensure that the large ratio of the observed speedup is attributable to the proposed method rather than changes in data type, we implemented the same hand-vectorized method using an 8-bit data type (handv-int8), dismissing overflow issues in the handv-int8 implementation.
It's important to note that, under operation requiring typical accuracy, an 8-bit multiplication yields a 16-bit result, and accumulation requires a 32-bit length. The handv-int8 implementation in this study deviates from this standard by not including extra instruction like $"reinterpret"$ to evaluate the efficiency of the CAMP micro-architecture beyond the data-type converting.
Consequently, while the handv-int8  method benefits from disregarding overflow and utilizes an 8-bit accumulator—which can lead to erroneous results—our proposed CAMP still significantly outperforms this approach. Experimental results in Figure~\ref{fig:cnn_speedup} reveals that the handv-int8 operations achieve a $2.5\times$ speedup over the 32-bit counterparts for GeMM operation in average, while CAMP4-bit yields $7\times$ greater efficiency compared to 32-bit operations. Despite overlooking certain overheads, we can attribute a significant portion of this speedup, specifically $4.5\times$, to the advantages offered by the CAMP.


\textbf{Significant Reduction in Instruction Count}:
The instruction count results in Figure~\ref{fig:cnn_speedup} and Figure~\ref{fig:llm_speedup}, validate the findings related to clock cycles, further underscoring the efficiency of the proposed architecture. As illustrated by the results, CAMP significantly reduces the number of required instructions, even when compared to aggressively optimized libraries like OpenBLAS.

\textbf{CAMP Alleviates Functional Unit Burden and stalls}:

According to Figure~\ref{fig:camp_busy_rate}, there is a significant reduction in the busy rate of functional unit for the CAMP architecture compared to the handv-int8 functional unit depicted in Figure~\ref{fig:functional_bar_busy_rate}. This reduction highlights the efficiency of the CAMP architecture in alleviating the load on the functional unit. Furthermore, the breakdown of stalls into Functional Unit, Read, and Write categories provide deeper insights into the operational delays, where a substantial portion of stalls are associated with storing operations, emphasizing that the proposed CAMP architecture is leveraging functional operations to enhance performance.

\textbf{Alignment of Vector Instruction Reduction with CAMP Expectations}:
By comparing the reduction in vector instructions (as shown in the heatmap) with the overall instruction reduction (scalar + vector) presented in Figure~\ref{fig:cnn_speedup} and Figure~\ref{fig:llm_speedup}, we observe that while total instruction count is reduced by approximately 50\% compared to OpenBLAS, the reduction in vector instructions alone is substantially greater. This highlights CAMP's ability to more efficiently utilize vector lanes and reduce redundant vector operations.
As elaborated in Section \ref{sec.cartesian}, the CAMP architecture not only enhances operations through additional functional units but also significantly reduces the number of vector load and store operations due to its efficiency in calculating outer products.
This reduction can extend to higher values due to the limited number of architectural vector registers, necessitating additional store and load operations to manage this limitation. Additionally, the CAMP architecture processes matrix multiplication with only one instruction, whereas conventional VAs in the investigated architecture requires multiple multiply-Add instructions. Consequently, given that the conventional method requires more operations like duplication or broadcasting, the reduction in instructions observed in the heatmap aligns with our expectations, showing more than an eightfold reduction.

\section{Related Work} \label{Sec:Related_work}

Matrix multiplication is a core operation that depends heavily on both hardware and software. On the software side, widely-used linear algebra libraries such as OpenBLAS~\cite{openblas}, BLIS~\cite{BLIS}, and ARM Performance Libraries (ARMPL)~\cite{armpl} have been evaluated across various architectures~\cite{valencia1,valencia2}. CAMP architecture is compatible with any VA configuration and we integrate it with the ulmBLAS library~\cite{ulmBLAS2021}.
We categorize hardware-relevant efforts into two sections: the first focuses on low-power SIMD-based designs, and the second surveys commercial vector architectures with native matrix multiplication support.

\subsection{SIMD-Support Processor}

Several studies have focused on enhancing edge devices by adding custom instructions and functional units to support lower bit-width GeMM operations. A detailed comparison for
the most relevant related works is then presented in Table~\ref{tbl:related}. The results presented in Table~\ref{tbl:related} are derived from the referenced papers, where the targeted benchmark focuses on the convolution operation. Specifically, the input tensor has dimensions \( H \times W \times F = 16 \times 16 \times 32 \), and the filter dimensions are \( F_{\text{out}} \times K_{\text{dim}} \times K_{\text{dim}} \times F_{\text{in}} = 64 \times 3 \times 3 \times 32 \).

PULP-NN\cite{pulp} is a well-known study that has inspired numerous subsequent works. This study is based on the PULP cluster of RISC-V-based processors, featuring an extended ISA to support narrow bit-width operations. The smallest supported data type is 8-bit, with packing and unpacking operations used to support 4-bit and 2-bit sub-byte data. However, the overhead of additional instructions required for casting 2-bit and 4-bit operands to 8-bit operands results in a degradation of speedup compared to RV32IMC. Specifically, the speedup drops from 8.8× for 8-bit operations to 3.69× and 4.22× for 4-bit and 2-bit data, respectively. By adding custom instructions and FUs to support 4-bit and 2-bit operations, the performance of the RISC-V core is enhanced for GeMM computations, as demonstrated in \cite{x_pulp}.
In addition to PULP-NN, some studies have proposed architectures to support mixed-precision data types. By extending PULP-NN, mixed-precision support has been added to the RISC-V core processor\cite{pulp_mix}. However, similar to PULP-NN, it suffers from the overhead associated with data-type casting.

To address this issue, \cite{ottavi2020mixed} proposes a RISC-V core with 2-bit and 4-bit MAC units along with the related ISA, effectively reducing the overhead associated with casting.
Additionally, based on the binary segmentation principle, low bit-width operations are proposed in \cite{enrico}, and by adding a software library and instructions, mixed-precision support for the architecture is achieved in \cite{Enrico_HPCA}.

According to the results in Table~\ref{tbl:related}, the proposed CAMP implementation, achieves up to 3× higher throughput for 8-bit data compared to best state-of-the-art. For 4-bit data, which represents the edge of reasonable accuracy in CNNs, it delivers up to 3× the throughput of the best 2-bit implementations reported in the literature, while maintaining power efficiency within acceptable limits. The simplicity of the proposed method eliminates the need for additional instructions to pack or unpack data~\cite{pulp,pulp_mix,ottavi2020mixed,x_pulp} or extra control unit overhead as observed in prior works~\cite{Enrico_HPCA}. Compared to ~\cite{Enrico_HPCA}, 
our design delivers a similar order of magnitude power efficiency while achieving significantly higher throughput, reaching up to 21.7 GOPS. The evaluation confirm that our architecture is both high-performance and energy-efficient. It also scales effectively across 8-bit and 4-bit configurations, demonstrating strong throughput and resource efficiency.

\subsection{Vector Architecture}
In terms of the VAs, IBM Matrix-Multiply Assist (MMA)~\cite{IBM}, Intel VNNI\cite{intel2023}, \cite{IntelDLBoost2023}, Arm MMLA~\cite{ARM_MMLA} implement instructions on CPU for matrix multiplication operation. 
The Intel Xeon processor with AVX512 VNNI speedups are 1.8x and 1.5x for Resnet-50 and, MobileNet-SSD, respectively compared to the AVX512 without VNNI~\cite{Intel_speedup}.
ARM introduced the MMLA instruction in ARMv8.6 for integer matrix multiplication, operating on 128-bit quadwords. It multiplies a \(2 \times 8\) row-major matrix with a \(2 \times 8\) column-major matrix and accumulates into a \(2 \times 2\) row-major matrix. This layout conflicts with the GotoBLAS algorithm, which expects column-major inputs and outputs. By modifying the packing strategy in our hand-vectorized implementation, we successfully integrated and validated MMLA support.

\begin{figure}[t]
  \centering
  \includegraphics[width=.47\textwidth]{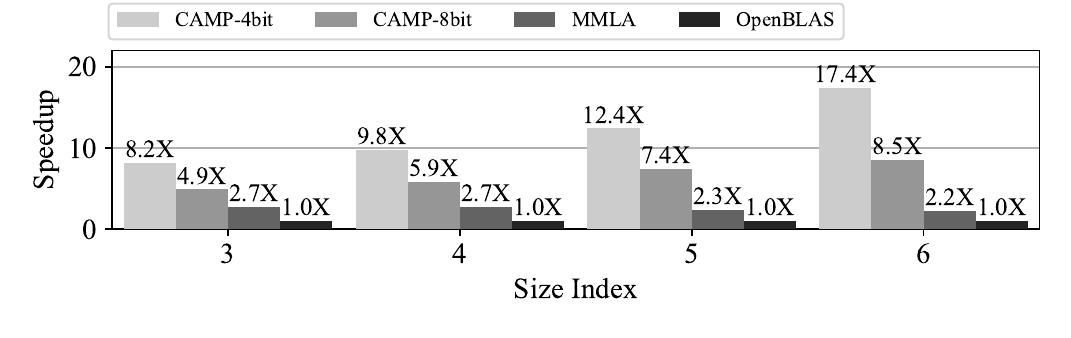} 
 
  \caption{
  Matrix multiplication performance comparison across different matrix sizes, with CAMP, and MMLA instructions. Results are normalized to OpenBLAS performance as the baseline.}

  \label{fig:MMLA_vs_CAMP}

\end{figure}

To evaluate the performance of the MMLA instruction and VNNI instruction to compare it with the proposed CAMP architecture, we run matrix multiplication operations for four different square matrices on an Yitian 710 core~\cite{alibaba2022server}.
We selected OpenBLAS as a baseline for performance comparison. Accordingly, 32-bit OpenBLAS sgemm was compiled and run for ARMv8-SVE 
on gem5, ARMv8.6-SVE2 on the Yitian as a baseline. Then, matrix multiplication with CAMP, and MMLA instruction was performed on the gem5 simulator, and Yitan, respectively. The comparison results are presented in Fig.~\ref{fig:MMLA_vs_CAMP}.
The proposed CAMP method outperform MMLA instructions. According to the experimental results, the performance of CAMP becomes more significant with the increase in the size of the matrix in the multiplication operation.
Furthermore, the algorithm for matrix multiplication based on the MMLA instruction heavily relies on the number of registers, and by increasing the block size, the speedup degrades compared to OpenBLAS, which aligns with the findings in~\cite{Yitian710}.     

\begin{table}[]
\caption{Performance and Efficiency Comparison with State-of-the-Art Methods}

\resizebox{\columnwidth}{!}{%
\begin{tabular}{ccccccccc}

\cline{ 1-9}
 & \multicolumn{5}{c}{\multirow{2}{*}{Architecture}} & \multicolumn{3}{c}{Benchmark} \\ \cline{7-8} 
     & \multicolumn{5}{c}{} & \multicolumn{3}{c}{Convolution} \\ \cline{2-6} \cline{7-8} 
 & \multirow{2}{*}{Data Size} & \multirow{2}{*}{SoC} & Freq & Tech & Area & Perf. & Energy Eff. \\
 &  &  & {[}GHz{]} & {[}$nm${]} & {[}$mm^2${]} & {[}GOPS{]} & {[}TOPS/W{]} &  \\ \hline
\cite{pulp} & 8b/4b/2b & RV32 & 0.17 & - & - & 0.6-0.2 & - & \\
\cite{pulp_mix} & 8b/4b/2b & 8$\times$ RV32 & 0.17 & - & - & 6.1-2.4 & - & \\
\cite{ottavi2020mixed} & 8b/4b/2b & RV32 & 0.25 & 22 & 0.002 & 1.1-3.3 & 0.2-0.6 &  \\
\cite{x_pulp} & 8b/4b/2b & 8 $\times$ RV32 & 0.6 & 22 & 8 $\times$ 0.04 & 19.8-47.9 & 0.7-1.1 &  \\
\cite{Enrico_HPCA} & All 8b-2b & RV64 & 1.2 & 22 & 0.0136 & 4.2-7.9 & 0.4-0.8 &  \\
This work & 8b/4b & RV64 & 1 & 22 & 0.0782 & 12.6-21.7 & 0.2-0.3 & 
\end{tabular}%
}
\label{tbl:related}

\end{table}

\section{Conclusion} \label{Sec:Conclusion}
\textls[-25]{
This paper presented the CAMP architecture, a co-designed hardware-software solution for accelerating matrix multiplication in VAs and SIMD-support units. Aligning with high performance libraries, CAMP enables efficient execution of quantized matrix multiplications without disrupting the standard vector processing pipeline. The architecture supports both 8-bit and 4-bit data types, achieving up to 17$\times$ performance improvement and over 80\% energy reduction compared to optimized software libraries running on commercial vector processors.
Its performance is validated across a wide set of LLM and CNN models, highlighting its generality and robustness.

CAMP paves the way for future vector extensions that natively support low-precision workloads, making it a compelling direction for next-generation edge and datacenter compute platforms.
}

\bibliographystyle{ACM-Reference-Format}
\bibliography{sample-base}


\begin{thebibliography}{67}


\ifx \showCODEN    \undefined \def \showCODEN     #1{\unskip}     \fi
\ifx \showISBNx    \undefined \def \showISBNx     #1{\unskip}     \fi
\ifx \showISBNxiii \undefined \def \showISBNxiii  #1{\unskip}     \fi
\ifx \showISSN     \undefined \def \showISSN      #1{\unskip}     \fi
\ifx \showLCCN     \undefined \def \showLCCN      #1{\unskip}     \fi
\ifx \shownote     \undefined \def \shownote      #1{#1}          \fi
\ifx \showarticletitle \undefined \def \showarticletitle #1{#1}   \fi
\ifx \showURL      \undefined \def \showURL       {\relax}        \fi
\providecommand\bibfield[2]{#2}
\providecommand\bibinfo[2]{#2}
\providecommand\natexlab[1]{#1}
\providecommand\showeprint[2][]{arXiv:#2}

\bibitem[AMD(2023)]%
        {AMDCDNA32023}
 \bibinfo{year}{2023}\natexlab{}.
\newblock \bibinfo{title}{AMD CDNA 3 Architecture}.
\newblock \bibinfo{howpublished}{\url{https://www.amd.com/content/dam/amd/en/documents/instinct-tech-docs/white-papers/amd-cdna-3-white-paper.pdf}}.
\newblock
\newblock
\shownote{[Web, accessed \today]}.


\bibitem[Nvi(2023)]%
        {NvidiaAmpere2023}
 \bibinfo{year}{2023}\natexlab{}.
\newblock \bibinfo{title}{NVIDIA Ampere GA102 GPU Architecture}.
\newblock \bibinfo{howpublished}{\url{https://www.nvidia.com/content/PDF/nvidia-ampere-ga-102-gpu-architecture-whitepaper-v2.pdf}}.
\newblock
\newblock
\shownote{[Web, accessed \today]}.


\bibitem[ris(2024)]%
        {riscv_gnu_toolchain}
 \bibinfo{year}{2024}\natexlab{}.
\newblock \bibinfo{title}{RISC-V GNU Compiler Toolchain}.
\newblock \bibinfo{howpublished}{\textit{Online}}.
\newblock
\newblock
\shownote{Available: \url{https://github.com/riscv/riscv-gnu-toolchain}}.


\bibitem[Abadi et~al\mbox{.}(2016)]%
        {tensorflow}
\bibfield{author}{\bibinfo{person}{Mart{\'\i}n Abadi}, \bibinfo{person}{Ashish Agarwal}, \bibinfo{person}{Paul Barham}, \bibinfo{person}{Eugene Brevdo}, \bibinfo{person}{Zhifeng Chen}, \bibinfo{person}{Craig Citro}, \bibinfo{person}{Greg~S Corrado}, \bibinfo{person}{Andy Davis}, \bibinfo{person}{Jeffrey Dean}, \bibinfo{person}{Matthieu Devin}, {et~al\mbox{.}}} \bibinfo{year}{2016}\natexlab{}.
\newblock \showarticletitle{Tensorflow: Large-scale machine learning on heterogeneous distributed systems}.
\newblock \bibinfo{journal}{\emph{arXiv preprint arXiv:1603.04467}} (\bibinfo{year}{2016}).
\newblock


\bibitem[Alaejos et~al\mbox{.}(2023)]%
        {valencia1}
\bibfield{author}{\bibinfo{person}{Guillermo Alaejos}, \bibinfo{person}{Adri{\'a}n Castell{\'o}}, \bibinfo{person}{H{\'e}ctor Mart{\'\i}nez}, \bibinfo{person}{Pedro Alonso-Jord{\'a}}, \bibinfo{person}{Francisco~D Igual}, {and} \bibinfo{person}{Enrique~S Quintana-Ort{\'\i}}.} \bibinfo{year}{2023}\natexlab{}.
\newblock \showarticletitle{Micro-kernels for portable and efficient matrix multiplication in deep learning}.
\newblock \bibinfo{journal}{\emph{The Journal of Supercomputing}} (\bibinfo{year}{2023}).
\newblock


\bibitem[{Alibaba Cloud}(2022)]%
        {alibaba2022server}
\bibfield{author}{\bibinfo{person}{{Alibaba Cloud}}.} \bibinfo{year}{2022}\natexlab{}.
\newblock \bibinfo{title}{{Alibaba Cloud Unveils New Server Chips to Optimize Cloud Computing Services}}.
\newblock
\newblock
\shownote{[Online]. Available: \url{https://www.alibabacloud.com/blog/598159}}.


\bibitem[{Amazon Web Services}(2023)]%
        {aws_inferentia2}
\bibfield{author}{\bibinfo{person}{{Amazon Web Services}}.} \bibinfo{year}{2023}\natexlab{}.
\newblock \bibinfo{title}{{AWS Inferentia}}.
\newblock \bibinfo{howpublished}{\url{https://aws.amazon.com/machine-learning/inferentia/}}.
\newblock
\newblock
\shownote{[Web, accessed \today]]}.


\bibitem[{Apple Inc.}(2023)]%
        {apple2023m3}
\bibfield{author}{\bibinfo{person}{{Apple Inc.}}} \bibinfo{year}{2023}\natexlab{}.
\newblock \bibinfo{title}{{Apple unveils M3, M3 Pro, and M3 Max, the most advanced chips for a personal computer}}.
\newblock \bibinfo{howpublished}{\url{https://www.apple.com/newsroom/2023/10/apple-unveils-m3-m3-pro-and-m3-max-the-most-advanced-chips-for-a-personal-computer/}}.
\newblock


\bibitem[Arima et~al\mbox{.}(2021)]%
        {arima2021power}
\bibfield{author}{\bibinfo{person}{Eishi Arima}, \bibinfo{person}{Yuetsu Kodama}, \bibinfo{person}{Tetsuya Odajima}, \bibinfo{person}{Miwako Tsuji}, {and} \bibinfo{person}{Mitsuhisa Sato}.} \bibinfo{year}{2021}\natexlab{}.
\newblock \showarticletitle{Power/performance/area evaluations for next-generation hpc processors using the a64fx chip}. In \bibinfo{booktitle}{\emph{2021 IEEE Symposium in Low-Power and High-Speed Chips (COOL CHIPS)}}. IEEE.
\newblock


\bibitem[ARM(Year)]%
        {armpl}
\bibfield{author}{\bibinfo{person}{ARM}.} \bibinfo{year}{Year}\natexlab{}.
\newblock \bibinfo{title}{{ARM Performance Libraries (ARMPL)}}.
\newblock \bibinfo{howpublished}{\url{https://developer.arm.com/tools-and-software/server-and-hpc/compile/arm-compiler-for-linux/arm-performance-libraries}}.
\newblock
\newblock
\shownote{Version x.x}.


\bibitem[Binkert et~al\mbox{.}(2011)]%
        {gem5}
\bibfield{author}{\bibinfo{person}{Nathan Binkert}, \bibinfo{person}{Bradford Beckmann}, \bibinfo{person}{Gabriel Black}, \bibinfo{person}{Steven~K Reinhardt}, \bibinfo{person}{Ali Saidi}, \bibinfo{person}{Arkaprava Basu}, \bibinfo{person}{Joel Hestness}, \bibinfo{person}{Derek~R Hower}, \bibinfo{person}{Tushar Krishna}, \bibinfo{person}{Somayeh Sardashti}, {et~al\mbox{.}}} \bibinfo{year}{2011}\natexlab{}.
\newblock \showarticletitle{The gem5 simulator}.
\newblock \bibinfo{journal}{\emph{ACM SIGARCH computer architecture news}} (\bibinfo{year}{2011}).
\newblock


\bibitem[Boutros et~al\mbox{.}(2020)]%
        {Intel_FPGA}
\bibfield{author}{\bibinfo{person}{Andrew Boutros}, \bibinfo{person}{Eriko Nurvitadhi}, \bibinfo{person}{Rui Ma}, \bibinfo{person}{Sergey Gribok}, \bibinfo{person}{Zhipeng Zhao}, \bibinfo{person}{James~C Hoe}, \bibinfo{person}{Vaughn Betz}, {and} \bibinfo{person}{Martin Langhammer}.} \bibinfo{year}{2020}\natexlab{}.
\newblock \showarticletitle{Beyond peak performance: Comparing the real performance of AI-optimized FPGAs and GPUs}. In \bibinfo{booktitle}{\emph{2020 international conference on field-programmable technology (ICFPT)}}. IEEE.
\newblock


\bibitem[Bruschi et~al\mbox{.}(2020)]%
        {pulp_mix}
\bibfield{author}{\bibinfo{person}{Nazareno Bruschi}, \bibinfo{person}{Angelo Garofalo}, \bibinfo{person}{Francesco Conti}, \bibinfo{person}{Giuseppe Tagliavini}, {and} \bibinfo{person}{Davide Rossi}.} \bibinfo{year}{2020}\natexlab{}.
\newblock \showarticletitle{Enabling mixed-precision quantized neural networks in extreme-edge devices}. In \bibinfo{booktitle}{\emph{Proceedings of the 17th ACM International Conference on Computing Frontiers}}. \bibinfo{pages}{217--220}.
\newblock


\bibitem[Burrello et~al\mbox{.}(2021)]%
        {dory}
\bibfield{author}{\bibinfo{person}{Alessio Burrello}, \bibinfo{person}{Angelo Garofalo}, \bibinfo{person}{Nazareno Bruschi}, \bibinfo{person}{Giuseppe Tagliavini}, \bibinfo{person}{Davide Rossi}, {and} \bibinfo{person}{Francesco Conti}.} \bibinfo{year}{2021}\natexlab{}.
\newblock \showarticletitle{Dory: Automatic end-to-end deployment of real-world dnns on low-cost iot mcus}.
\newblock \bibinfo{journal}{\emph{IEEE Trans. Comput.}} \bibinfo{volume}{70}, \bibinfo{number}{8} (\bibinfo{year}{2021}), \bibinfo{pages}{1253--1268}.
\newblock


\bibitem[Cai et~al\mbox{.}(2023)]%
        {Micro_accelerator}
\bibfield{author}{\bibinfo{person}{Xuyi Cai}, \bibinfo{person}{Ying Wang}, \bibinfo{person}{Xiaohan Ma}, \bibinfo{person}{Yinhe Han}, {and} \bibinfo{person}{Lei Zhang}.} \bibinfo{year}{2023}\natexlab{}.
\newblock \showarticletitle{DeepBurning-SEG: Generating DNN Accelerators of Segment-Grained Pipeline Architecture}. In \bibinfo{booktitle}{\emph{Proceedings of the 55th Annual IEEE/ACM International Symposium on Microarchitecture(MICRO)}}.
\newblock


\bibitem[Capotondi et~al\mbox{.}(2020)]%
        {cmix}
\bibfield{author}{\bibinfo{person}{Alessandro Capotondi}, \bibinfo{person}{Manuele Rusci}, \bibinfo{person}{Marco Fariselli}, {and} \bibinfo{person}{Luca Benini}.} \bibinfo{year}{2020}\natexlab{}.
\newblock \showarticletitle{CMix-NN: Mixed low-precision CNN library for memory-constrained edge devices}.
\newblock \bibinfo{journal}{\emph{IEEE Transactions on Circuits and Systems II: Express Briefs}} \bibinfo{volume}{67}, \bibinfo{number}{5} (\bibinfo{year}{2020}), \bibinfo{pages}{871--875}.
\newblock


\bibitem[Castell{\'o} et~al\mbox{.}(2022)]%
        {valencia2}
\bibfield{author}{\bibinfo{person}{Adri{\'a}n Castell{\'o}}, \bibinfo{person}{Enrique~S Quintana-Ort{\'\i}}, {and} \bibinfo{person}{Francisco~D Igual}.} \bibinfo{year}{2022}\natexlab{}.
\newblock \showarticletitle{Anatomy of the BLIS family of algorithms for matrix multiplication}. In \bibinfo{booktitle}{\emph{2022 30th Euromicro International Conference on Parallel, Distributed and Network-based Processing (PDP)}}. IEEE, \bibinfo{pages}{92--99}.
\newblock


\bibitem[Chellapilla et~al\mbox{.}(2006)]%
        {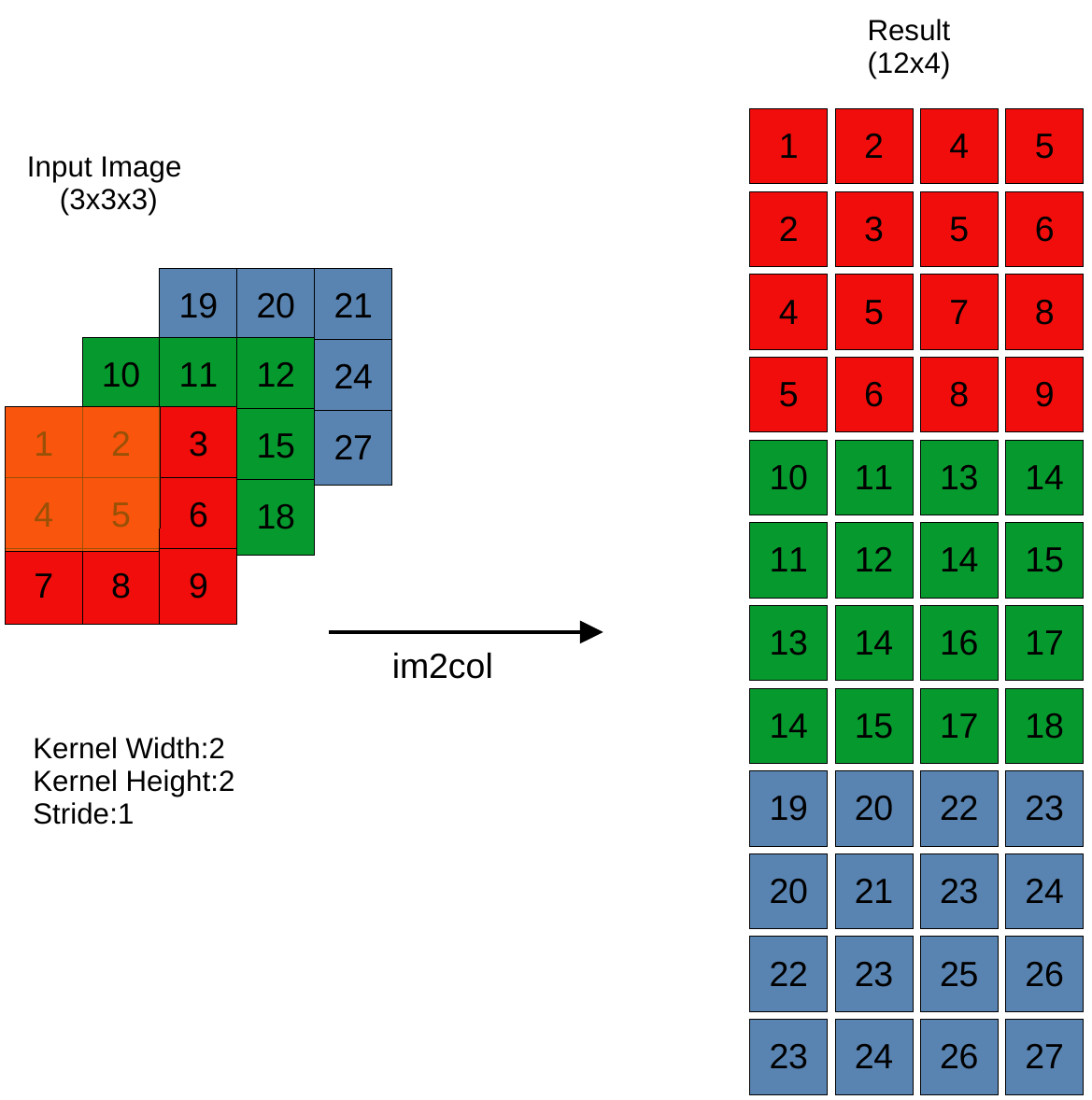}
\bibfield{author}{\bibinfo{person}{Kumar Chellapilla}, \bibinfo{person}{Sidd Puri}, {and} \bibinfo{person}{Patrice Simard}.} \bibinfo{year}{2006}\natexlab{}.
\newblock \showarticletitle{High performance convolutional neural networks for document processing}. In \bibinfo{booktitle}{\emph{Tenth international workshop on frontiers in handwriting recognition}}. Suvisoft.
\newblock


\bibitem[Chen et~al\mbox{.}(2014)]%
        {dadiannao}
\bibfield{author}{\bibinfo{person}{Yunji Chen}, \bibinfo{person}{Tao Luo}, \bibinfo{person}{Shaoli Liu}, \bibinfo{person}{Shijin Zhang}, \bibinfo{person}{Liqiang He}, \bibinfo{person}{Jia Wang}, \bibinfo{person}{Ling Li}, \bibinfo{person}{Tianshi Chen}, \bibinfo{person}{Zhiwei Xu}, \bibinfo{person}{Ninghui Sun}, {et~al\mbox{.}}} \bibinfo{year}{2014}\natexlab{}.
\newblock \showarticletitle{Dadiannao: A machine-learning supercomputer}. In \bibinfo{booktitle}{\emph{Proceedings of the 47th Annual IEEE/ACM International Symposium on Microarchitecture(MICRO)}}.
\newblock


\bibitem[Corporation(2023)]%
        {IntelDLBoost2023}
\bibfield{author}{\bibinfo{person}{Intel Corporation}.} \bibinfo{year}{2023}\natexlab{}.
\newblock \bibinfo{title}{Intel DL Boost (VNNI). Low Precision Integer Operations}.
\newblock \bibinfo{howpublished}{\url{https://ai.intel.com/intel-deep-learning-boost}}.
\newblock
\newblock
\shownote{[Web, accessed \today]}.


\bibitem[Corporation(2024)]%
        {nvidia2024blackwell}
\bibfield{author}{\bibinfo{person}{NVIDIA Corporation}.} \bibinfo{year}{2024}\natexlab{}.
\newblock \bibinfo{booktitle}{\emph{NVIDIA Blackwell Architecture Technical Overview}}.
\newblock
\urldef\tempurl%
\url{https://resources.nvidia.com/en-us-blackwell-architecture}
\showURL{%
\tempurl}
\newblock
\shownote{[Web, accessed \today]}.


\bibitem[Davoodi et~al\mbox{.}(2019)]%
        {NVIDIA}
\bibfield{author}{\bibinfo{person}{Pooya Davoodi}, \bibinfo{person}{Guangda Lai}, \bibinfo{person}{Trevor Morris}, {and} \bibinfo{person}{Siddharth Sharma}.} \bibinfo{year}{2019}\natexlab{}.
\newblock \bibinfo{title}{High performance inference with TensorRT Integration}.
\newblock \bibinfo{howpublished}{TensorFlow Blog}.
\newblock
\newblock
\shownote{\url{https://blog.tensorflow.org/2019/06/high-performance-inference-with-TensorRT.html}}.


\bibitem[{Development Team at University of Ulm}(2021)]%
        {ulmBLAS2021}
\bibfield{author}{\bibinfo{person}{{Development Team at University of Ulm}}.} \bibinfo{year}{2021}\natexlab{}.
\newblock \bibinfo{booktitle}{\emph{{ulmBLAS User Manual}}}.
\newblock University of Ulm.
\newblock


\bibitem[Dukhan et~al\mbox{.}(2018)]%
        {Facebook}
\bibfield{author}{\bibinfo{person}{Marat Dukhan}, \bibinfo{person}{Yiming Wu}, {and} \bibinfo{person}{Hao Lu}.} \bibinfo{year}{2018}\natexlab{}.
\newblock \bibinfo{title}{QNNPACK: Open source library for optimized mobile deep learning}.
\newblock


\bibitem[Garofalo et~al\mbox{.}(2020)]%
        {pulp}
\bibfield{author}{\bibinfo{person}{Angelo Garofalo}, \bibinfo{person}{Manuele Rusci}, \bibinfo{person}{Francesco Conti}, \bibinfo{person}{Davide Rossi}, {and} \bibinfo{person}{Luca Benini}.} \bibinfo{year}{2020}\natexlab{}.
\newblock \showarticletitle{PULP-NN: Accelerating quantized neural networks on parallel ultra-low-power RISC-V processors}.
\newblock \bibinfo{journal}{\emph{Philosophical Transactions of the Royal Society A}} \bibinfo{volume}{378}, \bibinfo{number}{2164} (\bibinfo{year}{2020}), \bibinfo{pages}{20190155}.
\newblock


\bibitem[Garofalo et~al\mbox{.}(2021)]%
        {x_pulp}
\bibfield{author}{\bibinfo{person}{Angelo Garofalo}, \bibinfo{person}{Giuseppe Tagliavini}, \bibinfo{person}{Francesco Conti}, \bibinfo{person}{Luca Benini}, {and} \bibinfo{person}{Davide Rossi}.} \bibinfo{year}{2021}\natexlab{}.
\newblock \showarticletitle{Xpulpnn: Enabling energy efficient and flexible inference of quantized neural networks on risc-v based iot end nodes}.
\newblock \bibinfo{journal}{\emph{IEEE Transactions on Emerging Topics in Computing}} \bibinfo{volume}{9}, \bibinfo{number}{3} (\bibinfo{year}{2021}), \bibinfo{pages}{1489--1505}.
\newblock


\bibitem[Genc et~al\mbox{.}(2024)]%
        {Stellar_micro}
\bibfield{author}{\bibinfo{person}{Hasan~Nazim Genc}, \bibinfo{person}{Hansung Kim}, \bibinfo{person}{Prashanth Ganesh}, {and} \bibinfo{person}{Yakun~Sophia Shao}.} \bibinfo{year}{2024}\natexlab{}.
\newblock \showarticletitle{Stellar: An Automated Design Framework for Dense and Sparse Spatial Accelerators}. In \bibinfo{booktitle}{\emph{Proceedings of the 57th IEEE/ACM International Symposium on Microarchitecture (MICRO)}}.
\newblock


\bibitem[Georganas et~al\mbox{.}(2018)]%
        {intelMKL}
\bibfield{author}{\bibinfo{person}{Evangelos Georganas}, \bibinfo{person}{Sasikanth Avancha}, \bibinfo{person}{Kunal Banerjee}, \bibinfo{person}{Dhiraj Kalamkar}, \bibinfo{person}{Greg Henry}, \bibinfo{person}{Hans Pabst}, {and} \bibinfo{person}{Alexander Heinecke}.} \bibinfo{year}{2018}\natexlab{}.
\newblock \showarticletitle{Anatomy of high-performance deep learning convolutions on simd architectures}. In \bibinfo{booktitle}{\emph{SC18: International Conference for High Performance Computing, Networking, Storage and Analysis}}. IEEE.
\newblock


\bibitem[{Google Cloud}(2023)]%
        {google_cloud_tpu}
\bibfield{author}{\bibinfo{person}{{Google Cloud}}.} \bibinfo{year}{2023}\natexlab{}.
\newblock \bibinfo{title}{{System Architecture: TPU VM}}.
\newblock \bibinfo{howpublished}{\url{https://cloud.google.com/tpu/docs/system-architecture-tpu-vm}}.
\newblock


\bibitem[Goto and Geijn(2008)]%
        {Goto_anatomy}
\bibfield{author}{\bibinfo{person}{Kazushige Goto} {and} \bibinfo{person}{Robert A. van~de Geijn}.} \bibinfo{year}{2008}\natexlab{}.
\newblock \showarticletitle{Anatomy of High-Performance Matrix Multiplication}.
\newblock \bibinfo{journal}{\emph{ACM Trans. Math. Softw.}} (\bibinfo{year}{2008}).
\newblock
\href{https://doi.org/10.1145/1356052.1356053}{doi:\nolinkurl{10.1145/1356052.1356053}}


\bibitem[Goto and Van De~Geijn(2008)]%
        {goto_high_performance}
\bibfield{author}{\bibinfo{person}{Kazushige Goto} {and} \bibinfo{person}{Robert Van De~Geijn}.} \bibinfo{year}{2008}\natexlab{}.
\newblock \showarticletitle{High-Performance Implementation of the Level-3 BLAS}.
\newblock \bibinfo{journal}{\emph{ACM Trans. Math. Softw.}} (\bibinfo{year}{2008}).
\newblock


\bibitem[Ham et~al\mbox{.}(2020)]%
        {ham20203}
\bibfield{author}{\bibinfo{person}{Tae~Jun Ham}, \bibinfo{person}{Sung~Jun Jung}, \bibinfo{person}{Seonghak Kim}, \bibinfo{person}{Young~H Oh}, \bibinfo{person}{Yeonhong Park}, \bibinfo{person}{Yoonho Song}, \bibinfo{person}{Jung-Hun Park}, \bibinfo{person}{Sanghee Lee}, \bibinfo{person}{Kyoung Park}, \bibinfo{person}{Jae~W Lee}, {et~al\mbox{.}}} \bibinfo{year}{2020}\natexlab{}.
\newblock \showarticletitle{A\^{} 3: Accelerating attention mechanisms in neural networks with approximation}. In \bibinfo{booktitle}{\emph{2020 IEEE International Symposium on High Performance Computer Architecture (HPCA)}}.
\newblock


\bibitem[Huang et~al\mbox{.}(2019)]%
        {ecnn}
\bibfield{author}{\bibinfo{person}{Chao-Tsung Huang}, \bibinfo{person}{Yu-Chun Ding}, \bibinfo{person}{Huan-Ching Wang}, \bibinfo{person}{Chi-Wen Weng}, \bibinfo{person}{Kai-Ping Lin}, \bibinfo{person}{Li-Wei Wang}, {and} \bibinfo{person}{Li-De Chen}.} \bibinfo{year}{2019}\natexlab{}.
\newblock \showarticletitle{ecnn: A block-based and highly-parallel cnn accelerator for edge inference}. In \bibinfo{booktitle}{\emph{Proceedings of the 52nd Annual IEEE/ACM International Symposium on Microarchitecture(MICRO)}}.
\newblock


\bibitem[{Intel Corporation}(2023a)]%
        {intel2023}
\bibfield{author}{\bibinfo{person}{{Intel Corporation}}.} \bibinfo{year}{2023}\natexlab{a}.
\newblock \bibinfo{title}{Intel Architecture Instruction Set Extensions and Future Features}.
\newblock


\bibitem[{Intel Corporation}(2023b)]%
        {IntelXeHPG2023}
\bibfield{author}{\bibinfo{person}{{Intel Corporation}}.} \bibinfo{year}{2023}\natexlab{b}.
\newblock \bibinfo{booktitle}{\emph{Introduction to the Xe HPG Architecture}}.
\newblock \bibinfo{type}{{T}echnical {R}eport}.
\newblock
\urldef\tempurl%
\url{https://cdrdv2-public.intel.com/758302/introduction-to-the-xe-hpg-architecture-white-paper.pdf}
\showURL{%
\tempurl}
\newblock
\shownote{[Web, accessed \today]}.


\bibitem[Jacob and Warden(2022)]%
        {gemmlowp}
\bibfield{author}{\bibinfo{person}{Benoit Jacob} {and} \bibinfo{person}{Pete Warden}.} \bibinfo{year}{2022}\natexlab{}.
\newblock \bibinfo{title}{gemmlowp: A small self-contained low-precision GEMM library}.
\newblock
\newblock
\shownote{\url{https://github.com/google/gemmlowp}}.


\bibitem[Karatsuba and Ofman(1962)]%
        {karatsuba}
\bibfield{author}{\bibinfo{person}{Anatolii~Alekseevich Karatsuba} {and} \bibinfo{person}{Yu~P Ofman}.} \bibinfo{year}{1962}\natexlab{}.
\newblock \showarticletitle{Multiplication of many-digital numbers by automatic computers}. In \bibinfo{booktitle}{\emph{Doklady Akademii Nauk}}, Vol.~\bibinfo{volume}{145}. Russian Academy of Sciences, \bibinfo{pages}{293--294}.
\newblock


\bibitem[Kim et~al\mbox{.}(2024)]%
        {samsung}
\bibfield{author}{\bibinfo{person}{Byeongho Kim}, \bibinfo{person}{Sanghoon Cha}, \bibinfo{person}{Sangsoo Park}, \bibinfo{person}{Jieun Lee}, \bibinfo{person}{Sukhan Lee}, \bibinfo{person}{Shin-haeng Kang}, \bibinfo{person}{Jinin So}, \bibinfo{person}{Kyungsoo Kim}, \bibinfo{person}{Jin Jung}, \bibinfo{person}{Jong-Geon Lee}, {et~al\mbox{.}}} \bibinfo{year}{2024}\natexlab{}.
\newblock \showarticletitle{The Breakthrough Memory Solutions for Improved Performance on LLM Inference}.
\newblock \bibinfo{journal}{\emph{IEEE Micro}} (\bibinfo{year}{2024}).
\newblock


\bibitem[Lawson et~al\mbox{.}(1979)]%
        {BLAS}
\bibfield{author}{\bibinfo{person}{C.~L. Lawson}, \bibinfo{person}{R.~J. Hanson}, \bibinfo{person}{D.~R. Kincaid}, {and} \bibinfo{person}{F.~T. Krogh}.} \bibinfo{year}{1979}\natexlab{}.
\newblock \showarticletitle{Basic Linear Algebra Subprograms for Fortran Usage}.
\newblock \bibinfo{journal}{\emph{ACM Trans. Math. Softw.}} (\bibinfo{year}{1979}).
\newblock


\bibitem[LeCun et~al\mbox{.}(2010)]%
        {lecun2010mnist}
\bibfield{author}{\bibinfo{person}{Yann LeCun}, \bibinfo{person}{Corinna Cortes}, \bibinfo{person}{Chris Burges}, {et~al\mbox{.}}} \bibinfo{year}{2010}\natexlab{}.
\newblock \bibinfo{title}{MNIST handwritten digit database}.
\newblock


\bibitem[Luo et~al\mbox{.}(2023)]%
        {AQ2_micro}
\bibfield{author}{\bibinfo{person}{Yukui Luo}, \bibinfo{person}{Nuo Xu}, \bibinfo{person}{Hongwu Peng}, \bibinfo{person}{Chenghong Wang}, \bibinfo{person}{Shijin Duan}, \bibinfo{person}{Kaleel Mahmood}, \bibinfo{person}{Wujie Wen}, \bibinfo{person}{Caiwen Ding}, {and} \bibinfo{person}{Xiaolin Xu}.} \bibinfo{year}{2023}\natexlab{}.
\newblock \showarticletitle{Aq2pnn: Enabling two-party privacy-preserving deep neural network inference with adaptive quantization}. In \bibinfo{booktitle}{\emph{Proceedings of the 56th Annual IEEE/ACM International Symposium on Microarchitecture(MICRO)}}.
\newblock


\bibitem[Moons et~al\mbox{.}(2017)]%
        {envision}
\bibfield{author}{\bibinfo{person}{Bert Moons}, \bibinfo{person}{Roel Uytterhoeven}, \bibinfo{person}{Wim Dehaene}, {and} \bibinfo{person}{Marian Verhelst}.} \bibinfo{year}{2017}\natexlab{}.
\newblock \showarticletitle{14.5 envision: A 0.26-to-10tops/w subword-parallel dynamic-voltage-accuracy-frequency-scalable convolutional neural network processor in 28nm fdsoi}. In \bibinfo{booktitle}{\emph{2017 IEEE International Solid-State Circuits Conference (ISSCC)}}. IEEE.
\newblock


\bibitem[Mosur(2022)]%
        {Intel_speedup}
\bibfield{author}{\bibinfo{person}{Praveen Mosur}.} \bibinfo{year}{2022}\natexlab{}.
\newblock \showarticletitle{Built for the Edge: The Next-Generation IntelXeon D 2700 \& 1700 processors}. In \bibinfo{booktitle}{\emph{2022 IEEE Hot Chips 34 Symposium (HCS)}}.
\newblock


\bibitem[Odajima et~al\mbox{.}(2020)]%
        {odajima2020preliminary}
\bibfield{author}{\bibinfo{person}{Tetsuya Odajima}, \bibinfo{person}{Yuetsu Kodama}, \bibinfo{person}{Miwako Tsuji}, \bibinfo{person}{Motohiko Matsuda}, \bibinfo{person}{Yutaka Maruyama}, {and} \bibinfo{person}{Mitsuhisa Sato}.} \bibinfo{year}{2020}\natexlab{}.
\newblock \showarticletitle{Preliminary performance evaluation of the Fujitsu A64FX using HPC applications}. In \bibinfo{booktitle}{\emph{2020 IEEE international conference on cluster computing (cluster)}}. IEEE.
\newblock


\bibitem[Okazaki et~al\mbox{.}(2020)]%
        {Fugako}
\bibfield{author}{\bibinfo{person}{Ryohei Okazaki}, \bibinfo{person}{Takekazu Tabata}, \bibinfo{person}{Sota Sakashita}, \bibinfo{person}{Kenichi Kitamura}, \bibinfo{person}{Noriko Takagi}, \bibinfo{person}{Hideki Sakata}, \bibinfo{person}{Takeshi Ishibashi}, \bibinfo{person}{Takeo Nakamura}, {and} \bibinfo{person}{Yuichiro Ajima}.} \bibinfo{year}{2020}\natexlab{}.
\newblock \showarticletitle{Supercomputer Fugaku Cpu A64fx realizing high performance, high-density packaging, and low power consumption}.
\newblock \bibinfo{journal}{\emph{Fujitsu Technical Review}} (\bibinfo{year}{2020}).
\newblock


\bibitem[Ottavi et~al\mbox{.}(2020)]%
        {ottavi2020mixed}
\bibfield{author}{\bibinfo{person}{Gianmarco Ottavi}, \bibinfo{person}{Angelo Garofalo}, \bibinfo{person}{Giuseppe Tagliavini}, \bibinfo{person}{Francesco Conti}, \bibinfo{person}{Luca Benini}, {and} \bibinfo{person}{Davide Rossi}.} \bibinfo{year}{2020}\natexlab{}.
\newblock \showarticletitle{A mixed-precision RISC-V processor for extreme-edge DNN inference}. In \bibinfo{booktitle}{\emph{2020 IEEE Computer Society Annual Symposium on VLSI (ISVLSI)}}. IEEE, \bibinfo{pages}{512--517}.
\newblock


\bibitem[Park et~al\mbox{.}(2022)]%
        {Diva_micro}
\bibfield{author}{\bibinfo{person}{Beomsik Park}, \bibinfo{person}{Ranggi Hwang}, \bibinfo{person}{Dongho Yoon}, \bibinfo{person}{Yoonhyuk Choi}, {and} \bibinfo{person}{Minsoo Rhu}.} \bibinfo{year}{2022}\natexlab{}.
\newblock \showarticletitle{Diva: An accelerator for differentially private machine learning}. In \bibinfo{booktitle}{\emph{Proceedings of the 55th IEEE/ACM International Symposium on Microarchitecture (MICRO)}}.
\newblock


\bibitem[Park et~al\mbox{.}(2018)]%
        {dnn_facebook}
\bibfield{author}{\bibinfo{person}{Jongsoo Park}, \bibinfo{person}{Maxim Naumov}, \bibinfo{person}{Protonu Basu}, \bibinfo{person}{Summer Deng}, \bibinfo{person}{Aravind Kalaiah}, \bibinfo{person}{Daya Khudia}, \bibinfo{person}{James Law}, \bibinfo{person}{Parth Malani}, \bibinfo{person}{Andrey Malevich}, \bibinfo{person}{Satish Nadathur}, {et~al\mbox{.}}} \bibinfo{year}{2018}\natexlab{}.
\newblock \showarticletitle{Deep learning inference in facebook data centers: Characterization, performance optimizations and hardware implications}.
\newblock \bibinfo{journal}{\emph{arXiv preprint arXiv:1811.09886}} (\bibinfo{year}{2018}).
\newblock


\bibitem[Rashidi et~al\mbox{.}(2023)]%
        {Unico_micro}
\bibfield{author}{\bibinfo{person}{Bahador Rashidi}, \bibinfo{person}{Chao Gao}, \bibinfo{person}{Shan Lu}, \bibinfo{person}{Zhisheng Wang}, \bibinfo{person}{Chunhua Zhou}, \bibinfo{person}{Di Niu}, {and} \bibinfo{person}{Fengyu Sun}.} \bibinfo{year}{2023}\natexlab{}.
\newblock \showarticletitle{Unico: Unified hardware software co-optimization for robust neural network acceleration}. In \bibinfo{booktitle}{\emph{Proceedings of the 56th Annual IEEE/ACM International Symposium on Microarchitecture(MICRO)}}.
\newblock


\bibitem[Reggiani et~al\mbox{.}(2022)]%
        {enrico}
\bibfield{author}{\bibinfo{person}{Enrico Reggiani}, \bibinfo{person}{Crist{\'o}bal~Ram{\'\i}rez Lazo}, \bibinfo{person}{Roger~Figueras Bagu{\'e}}, \bibinfo{person}{Adri{\'a}n Cristal}, \bibinfo{person}{Mauro Olivieri}, {and} \bibinfo{person}{Osman~Sabri Unsal}.} \bibinfo{year}{2022}\natexlab{}.
\newblock \showarticletitle{BiSon-e: a lightweight and high-performance accelerator for narrow integer linear algebra computing on the edge}. In \bibinfo{booktitle}{\emph{Proceedings of the 27th ACM International Conference on Architectural Support for Programming Languages and Operating Systems}}.
\newblock


\bibitem[Reggiani et~al\mbox{.}(2023)]%
        {Enrico_HPCA}
\bibfield{author}{\bibinfo{person}{Enrico Reggiani}, \bibinfo{person}{Alessandro Pappalardo}, \bibinfo{person}{Max Doblas}, \bibinfo{person}{Miquel Moreto}, \bibinfo{person}{Mauro Olivieri}, \bibinfo{person}{Osman~Sabri Unsal}, {and} \bibinfo{person}{Adrián Cristal}.} \bibinfo{year}{2023}\natexlab{}.
\newblock \showarticletitle{Mix-GEMM: An efficient HW-SW Architecture for Mixed-Precision Quantized Deep Neural Networks Inference on Edge Devices}. In \bibinfo{booktitle}{\emph{2023 IEEE International Symposium on High-Performance Computer Architecture (HPCA)}}.
\newblock


\bibitem[Remke and Breuer(2024)]%
        {remke2024hello}
\bibfield{author}{\bibinfo{person}{Stefan Remke} {and} \bibinfo{person}{Alexander Breuer}.} \bibinfo{year}{2024}\natexlab{}.
\newblock \showarticletitle{Hello SME! Generating Fast Matrix Multiplication Kernels Using the Scalable Matrix Extension}. In \bibinfo{booktitle}{\emph{SC24-W: Workshops of the International Conference for High Performance Computing, Networking, Storage and Analysis}}. IEEE, \bibinfo{pages}{1443--1454}.
\newblock


\bibitem[Sheng et~al\mbox{.}(2023)]%
        {large_language_model}
\bibfield{author}{\bibinfo{person}{Ying Sheng}, \bibinfo{person}{Lianmin Zheng}, \bibinfo{person}{Binhang Yuan}, \bibinfo{person}{Zhuohan Li}, \bibinfo{person}{Max Ryabinin}, \bibinfo{person}{Daniel~Y Fu}, \bibinfo{person}{Zhiqiang Xie}, \bibinfo{person}{Beidi Chen}, \bibinfo{person}{Clark Barrett}, \bibinfo{person}{Joseph~E Gonzalez}, {et~al\mbox{.}}} \bibinfo{year}{2023}\natexlab{}.
\newblock \showarticletitle{High-throughput generative inference of large language models with a single gpu}.
\newblock \bibinfo{journal}{\emph{arXiv preprint arXiv:2303.06865}} (\bibinfo{year}{2023}).
\newblock


\bibitem[Soria-Pardos et~al\mbox{.}(2022)]%
        {sargantana}
\bibfield{author}{\bibinfo{person}{V{\'\i}ctor Soria-Pardos}, \bibinfo{person}{Max Doblas}, \bibinfo{person}{Guillem L{\'o}pez-Parad{\'\i}s}, \bibinfo{person}{Gerard Cand{\'o}n}, \bibinfo{person}{Narc{\'\i}s Rodas}, \bibinfo{person}{Xavier Carril}, \bibinfo{person}{Pau Fontova-Must{\'e}}, \bibinfo{person}{Neiel Leyva}, \bibinfo{person}{Santiago Marco-Sola}, {and} \bibinfo{person}{Miquel Moret{\'o}}.} \bibinfo{year}{2022}\natexlab{}.
\newblock \showarticletitle{Sargantana: A 1 GHz+ in-order RISC-V processor with SIMD vector extensions in 22nm FD-SOI}. In \bibinfo{booktitle}{\emph{2022 25th Euromicro Conference on Digital System Design (DSD)}}. IEEE, \bibinfo{pages}{254--261}.
\newblock


\bibitem[Starke et~al\mbox{.}(2021)]%
        {IBM}
\bibfield{author}{\bibinfo{person}{William~J. Starke}, \bibinfo{person}{Brian~W. Thompto}, \bibinfo{person}{Jeff~A. Stuecheli}, {and} \bibinfo{person}{José~E. Moreira}.} \bibinfo{year}{2021}\natexlab{}.
\newblock \showarticletitle{IBM's POWER10 Processor}.
\newblock \bibinfo{journal}{\emph{IEEE Micro}} (\bibinfo{year}{2021}).
\newblock


\bibitem[Sung et~al\mbox{.}(2023)]%
        {madmac}
\bibfield{author}{\bibinfo{person}{Seunghwan Sung}, \bibinfo{person}{Sujin Hur}, \bibinfo{person}{Sungwoo Kim}, \bibinfo{person}{Dongho Ha}, \bibinfo{person}{Yunho Oh}, {and} \bibinfo{person}{Won~Woo Ro}.} \bibinfo{year}{2023}\natexlab{}.
\newblock \showarticletitle{Mad macce: Supporting multiply-add operations for democratizing matrix-multiplication accelerators}. In \bibinfo{booktitle}{\emph{Proceedings of the 56th Annual IEEE/ACM International Symposium on Microarchitecture(MICRO)}}.
\newblock


\bibitem[Synopsys({[n.\,d.]})]%
        {synopsys}
\bibfield{author}{\bibinfo{person}{Synopsys}.} \bibinfo{year}{[n.\,d.]}\natexlab{}.
\newblock \bibinfo{title}{{Synopsys}}.
\newblock \bibinfo{howpublished}{\url{https://www.synopsys.com/}}.
\newblock
\newblock
\shownote{[Web, accessed \today]}.


\bibitem[{Tesla, Inc.}(2023)]%
        {tesla2023dojowhitepaper}
\bibfield{author}{\bibinfo{person}{{Tesla, Inc.}}} \bibinfo{year}{2023}\natexlab{}.
\newblock \bibinfo{title}{{Tesla Dojo Technology}}.
\newblock \bibinfo{howpublished}{\url{https://digitalassets.tesla.com/tesla-contents/image/upload/tesla-dojo-technology.pdf}}.
\newblock
\newblock
\shownote{[[Web, accessed \today]]}.


\bibitem[Van~Zee and Smith(2017)]%
        {van2017implementing}
\bibfield{author}{\bibinfo{person}{Field~G Van~Zee} {and} \bibinfo{person}{Tyler~M Smith}.} \bibinfo{year}{2017}\natexlab{}.
\newblock \showarticletitle{Implementing high-performance complex matrix multiplication via the 3m and 4m methods}.
\newblock \bibinfo{journal}{\emph{ACM Transactions on Mathematical Software (TOMS)}} \bibinfo{volume}{44}, \bibinfo{number}{1} (\bibinfo{year}{2017}), \bibinfo{pages}{1--36}.
\newblock


\bibitem[Van~Zee and Van De~Geijn(2015)]%
        {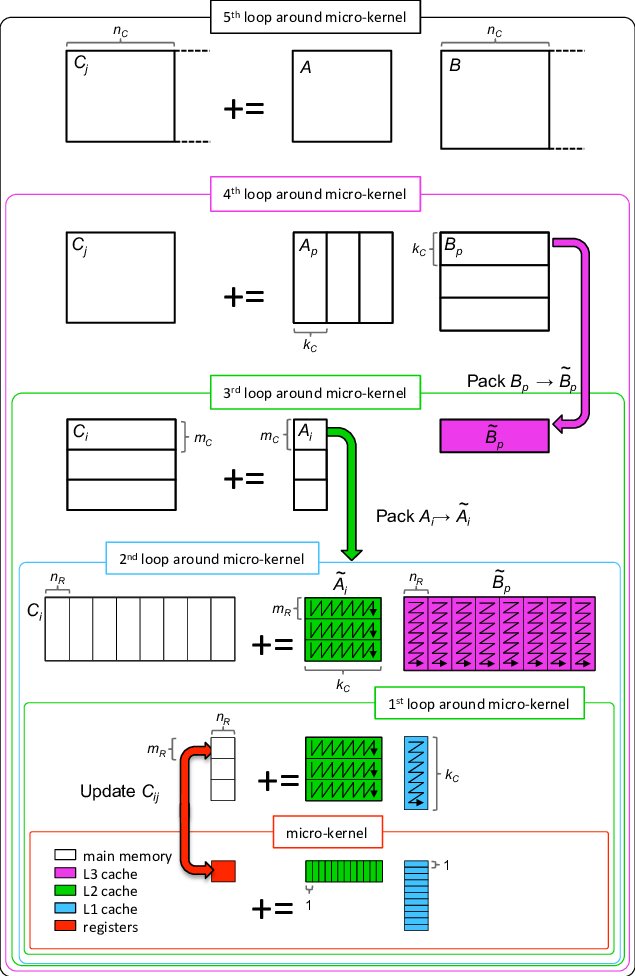}
\bibfield{author}{\bibinfo{person}{Field~G Van~Zee} {and} \bibinfo{person}{Robert~A Van De~Geijn}.} \bibinfo{year}{2015}\natexlab{}.
\newblock \showarticletitle{BLIS: A framework for rapidly instantiating BLAS functionality}.
\newblock \bibinfo{journal}{\emph{ACM Transactions on Mathematical Software (TOMS)}} (\bibinfo{year}{2015}).
\newblock


\bibitem[Vinyals and Le(2015)]%
        {chatbot}
\bibfield{author}{\bibinfo{person}{Oriol Vinyals} {and} \bibinfo{person}{Quoc Le}.} \bibinfo{year}{2015}\natexlab{}.
\newblock \showarticletitle{A neural conversational model}.
\newblock \bibinfo{journal}{\emph{arXiv preprint arXiv:1506.05869}} (\bibinfo{year}{2015}).
\newblock


\bibitem[Wang et~al\mbox{.}(2023)]%
        {nature}
\bibfield{author}{\bibinfo{person}{Hanchen Wang}, \bibinfo{person}{Tianfan Fu}, \bibinfo{person}{Yuanqi Du}, \bibinfo{person}{Wenhao Gao}, \bibinfo{person}{Kexin Huang}, \bibinfo{person}{Ziming Liu}, \bibinfo{person}{Payal Chandak}, \bibinfo{person}{Shengchao Liu}, \bibinfo{person}{Peter Van~Katwyk}, \bibinfo{person}{Andreea Deac}, {et~al\mbox{.}}} \bibinfo{year}{2023}\natexlab{}.
\newblock \showarticletitle{Scientific discovery in the age of artificial intelligence}.
\newblock \bibinfo{journal}{\emph{Nature}} (\bibinfo{year}{2023}).
\newblock


\bibitem[Wang et~al\mbox{.}(2021)]%
        {wang2021spatten}
\bibfield{author}{\bibinfo{person}{Hanrui Wang}, \bibinfo{person}{Zhekai Zhang}, {and} \bibinfo{person}{Song Han}.} \bibinfo{year}{2021}\natexlab{}.
\newblock \showarticletitle{Spatten: Efficient sparse attention architecture with cascade token and head pruning}. In \bibinfo{booktitle}{\emph{2021 IEEE International Symposium on High-Performance Computer Architecture (HPCA)}}.
\newblock


\bibitem[Weidmann(2021)]%
        {ARM_MMLA}
\bibfield{author}{\bibinfo{person}{Martin Weidmann}.} \bibinfo{year}{2021}\natexlab{}.
\newblock \bibinfo{booktitle}{\emph{Introducing the scalable matrix extension for the Armv9-A architecture}}.
\newblock \bibinfo{type}{{T}echnical {R}eport}. \bibinfo{institution}{Technical report}.
\newblock


\bibitem[Xianyi et~al\mbox{.}(2012)]%
        {openblas}
\bibfield{author}{\bibinfo{person}{Zhang Xianyi}, \bibinfo{person}{Wang Qian}, {and} \bibinfo{person}{Zaheer Chothia}.} \bibinfo{year}{2012}\natexlab{}.
\newblock \showarticletitle{OpenBLAS}.
\newblock \bibinfo{journal}{\emph{URL: http://xianyi. github. io/OpenBLAS}} (\bibinfo{year}{2012}).
\newblock


\bibitem[Xilinx({[n.\,d.]})]%
        {xilinx_alveo}
\bibfield{author}{\bibinfo{person}{Xilinx}.} \bibinfo{year}{[n.\,d.]}\natexlab{}.
\newblock \bibinfo{title}{Alveo U250 Data Center Accelerator Card}.
\newblock \bibinfo{howpublished}{\url{https://www.xilinx.com/products/boards-and-kits/alveo/u250.html}}.
\newblock
\newblock
\shownote{[Web, accessed \today]}.


\bibitem[Yu et~al\mbox{.}(2023)]%
        {Yitian710}
\bibfield{author}{\bibinfo{person}{Guosheng Yu}, \bibinfo{person}{Zhihong Lv}, \bibinfo{person}{Haijiang Wang}, \bibinfo{person}{Zilong Huang}, {and} \bibinfo{person}{Jicheng Chen}.} \bibinfo{year}{2023}\natexlab{}.
\newblock \showarticletitle{Task-aware Scheduling and Performance Optimization on Yitian710 SoC for GEMM-based Workloads on the Cloud}. In \bibinfo{booktitle}{\emph{2023 IEEE 5th International Conference on Artificial Intelligence Circuits and Systems (AICAS)}}.
\newblock


\end{thebibliography}

\end{document}